\documentclass[preprint]{aastex61}

\usepackage{amsmath}
\usepackage{graphicx}
\usepackage{natbib}
\usepackage{ulem}
\usepackage{bm}

\usepackage{xspace}
\usepackage{hyperref}

\usepackage{multirow}

\newtheorem{IVP}{IVP}
\newtheorem{EVP}{EVP}

\defcitealias{1996ApJ...472..398B}{BDBG96}
\defcitealias{1983SoPh...88..179E}{ER83}
\defcitealias{2015ApJ...810...87L}{LN15}
\defcitealias{2015ApJ...806...56O}{ORT15}

\newcommand\dispfrac[2]{\ensuremath{\displaystyle\frac{#1}{#2}}}
\newcommand\bigO{\ensuremath{\mathcal{O}}}
\def\myvect#1{\ensuremath{\bm{#1}}}

\newcommand{\Alf}{Alfv$\acute{\rm e}$n}

\newcommand{\Schrod}{Schr{\"o}dinger}

\newcommand{\kms}{\ensuremath{{\rm km~s}^{-1}}}

\newcommand{\dcl}{\ensuremath{{(d/R)}_{{\rm cri}, l}}}
\newcommand{\kappae}{\ensuremath{\kappa_{\rm e}}}
\newcommand{\ki}{\ensuremath{k_{\rm i}}}
\newcommand{\ke}{\ensuremath{k_{\rm e}}}
\newcommand{\kcut}{\ensuremath{k_{\rm cutoff}}}

\newcommand{\omgl}{\ensuremath{\omega_l}}
\newcommand{\omgcrit}{\ensuremath{\omega_{\rm crit}}} 

\newcommand{\Reff}{\ensuremath{R_{\rm eff}}}
\newcommand{\tauener}{\ensuremath{\tau_{\rm ener}}}
\newcommand{\taulongi}{\ensuremath{\tau_{\rm long}}}

\newcommand{\va}{\ensuremath{v_{\rm A}}}
\newcommand{\vai}{\ensuremath{v_{\rm Ai}}}
\newcommand{\vae}{\ensuremath{v_{\rm Ae}}}
\newcommand{\vhatd}{\ensuremath{\hat{v}^{(d)}}}

\newcommand{\rhoi}{\ensuremath{\rho_{\rm i}}}
\newcommand{\rhoe}{\ensuremath{\rho_{\rm e}}}

\newcommand{\vph}{\ensuremath{v_{\rm ph}}}

\submitjournal{ApJ}

\shorttitle{Standing Sausage Modes in Diffuse Coronal Loops}
\shortauthors{Li et al.}
\begin{document}

\title{Standing Sausage Perturbations in solar coronal loops with diffuse boundaries:
      An initial-value-problem perspective}

\correspondingauthor{Bo Li}
\email{bbl@sdu.edu.cn}

\author{Bo Li}
\affiliation{Shandong Provincial Key Laboratory of Optical Astronomy and Solar-Terrestrial Environment,
   Institute of Space Sciences, Shandong University, Weihai 264209, China}

\author{Shao-Xia Chen}
\affiliation{Shandong Provincial Key Laboratory of Optical Astronomy and Solar-Terrestrial Environment,
   Institute of Space Sciences, Shandong University, Weihai 264209, China}

\author{Ao-Long Li}
\affiliation{Shandong Provincial Key Laboratory of Optical Astronomy and Solar-Terrestrial Environment,
   Institute of Space Sciences, Shandong University, Weihai 264209, China}

\begin{abstract}
Working in pressureless magnetohydrodynamics,
    we examine the consequences of some peculiar dispersive properties of 
    linear fast sausage modes (FSMs) in one-dimensional cylindrical equilibria
    with a continuous radial density profile ($\rho_0(r)$).
As recognized recently on solid mathematical grounds, 
    cutoff axial wavenumbers may be absent for FSMs
    when $\rho_0(r)$ varies sufficiently slowly outside the nominal cylinder.
Trapped modes may therefore exist for arbitrary axial wavenumbers 
    and density contrasts, their axial phase speeds in the long-wavelength regime differing little from the external Alfv$\acute{\rm e}$n speed. 
If these trapped modes indeed show up in the solutions to the associated initial
    value problem (IVP), then FSMs have a much better chance to be observed
    than expected with classical theory, and can be invoked to account for
    a considerably broader range of periodicities than practiced. 
However, with axial fundamentals in active region loops as an example, 
    we show that this long-wavelength expectation is not seen in our finite-difference solutions to the IVP, the reason for which is then explored by 
    superposing the necessary eigenmodes to re-solve the IVP.
At least for the parameters we examine,     
    the eigenfunctions of trapped modes are characterized by a spatial extent 
    well exceeding the observationally reasonable range of the spatial extent of
    initial perturbations, meaning a negligible fraction of energy that
    a trapped mode can receive.
We conclude that the absence of cutoff wavenumbers for FSMs in the examined equilibrium
    does not guarantee a distinct temporal behavior.    
\end{abstract}


\section{INTRODUCTION}
\label{sec_intro}
{There have been abundant observational instances
    of low-frequency waves and oscillations in the structured solar corona}
    \citep[see, e.g., the reviews by][]
    {2007SoPh..246....3B, 2012RSPTA.370.3193D, 2016GMS...216..395W, 2016SSRv..200...75N}.
Combined with {magnetohydrodynamic (MHD)} wave theory, these observations 
    can help deduce the atmospheric parameters that prove difficult
    to directly measure, thereby constituting the field of
    ``coronal seismology''
    (see e.g., the reviews by 
    \citealt{2005LRSP....2....3N,2020ARA&A..58..441N};
    also the textbook by \citealt{2019CUP_Roberts}). 
Evidently, a physical interpretation needs to be assigned to    
    an observed oscillatory signal for it to be seismologically exploited.
For this purpose, it has been customary to contrast observations
    with the theoretical expectations for waves in field-aligned 
    cylinders that are structured only in the radial direction
    and in a step form 
    (developed by e.g., 
    \citealt[][hereafter \citetalias{1983SoPh...88..179E}]{1979A&A....76...20W,1982SoPh...75....3S,1983SoPh...88..179E};
    also
    	\citealt{1975IGAFS..37....3Z,1986SoPh..103..277C}).
It turns out that this apparently simple equilibrium supports
    a rich variety of waves, 
    and we restrict ourselves to the fast family
    \citep[see][for the most recent review on the slow family]          	{2021SSRv..217...34W}.
Indeed, radial fundamental kink modes in the sense of
	\citet{2009A&A...503..213G,2012ApJ...753..111G}    	 
    {have been amply identified and put to seismological practice
    \citep[see the review by][]{2021SSRv..217...73N}.
As an outcome, 
    the spatial variations of the magnetic field strength were deduced
    not only for individual active regions \citep[ARs,][]{2019ApJ...884L..40A}
    but also over a substantial fraction of the lower corona \citep{2020Sci...369..694Y}}.    
    
Candidate fast sausage modes (FSMs), however, have only been sporadically
    reported in coronal observations 
    \citep[see][for the most recent review]{2020SSRv..216..136L}.
As detailed therein, this rarity is intimately connected to the cutoff
    axial wavenumbers $\kcut$, 
    to explain which it suffices to consider the pressureless MHD.
In fact, we will adopt pressureless MHD throughout,
    and additionally restrict ourselves to flare loops and AR loops 
    as wave-guiding inhomogeneities
    \footnote{Sausage perturbations in flare current sheets have also been 
    invoked to account for, say, some fine structures in decimetric
    type IV radio bursts
    (\citealt{2011A&A...529A..96K,2012A&A...537A..46J};
    also \citealt{2020SSRv..216..136L} and references therein).
    We refrain from discussing such observations 
       to avoid the intricacies that cannot be addressed with pressureless MHD.
    In fact, we decide to leave out sausage modes
       in slab-type configurations altogether for the ease of description,
       even though they have been
       extensively examined \citep[e.g.,][]{1993SoPh..144..101M, 2004MNRAS.349..705N, 
       2013A&A...560A..97P, 2016ApJ...826...78Y, 2021MNRAS.505.3505K}.
    The approach we are to use, however, is sufficiently general.}.      
Let $R$ denote the cylinder radius, 
    and $\rhoi$ ($\rhoe$) the internal (external) density
    with $\rhoi > \rhoe$.
Likewise, let $\vai$ ($\vae$) represent the internal (external)
    \Alf\ speed. 
Standard analysis of the eigenvalue problem (EVP) on a laterally open domain
    then yields that FSMs in an \citetalias{1983SoPh...88..179E} 
    equilibrium possess a series of
    $k_{{\rm cutoff}, m} = j_{0, m}/(R \sqrt{\rhoi/\rhoe-1})$, where $j_{0, m}$
    is the $m$-th zero of Bessel $J_0$ with $m=1, 2, \cdots$
    \citep[e.g.,][]{1984ApJ...279..857R,2014ApJ...781...92V}.   
Now consider an open system unbounded in both the axial ($z$) 
    and radial ($r$) directions, and suppose that a cylinder
    is perturbed by an axisymmetric perturbation localized
    both radially and axially. 
The pertinent two-dimensional initial value problem (2D IVP) 
    has been examined rather extensively, with the majority of solutions
    found by directly evolving the MHD equations
    \citep[e.g.,][]{2015ApJ...814..135S,2016ApJ...833...51Y,2017ApJ...836....1Y}.
Physical insights, on the other hand, can also be gleaned 
    from a modal approach, which was discussed heuristically
    by \citet{1986NASCP2449..347E} and made more formal by 
    \citet{1996ApJ...472..398B}.
Our study makes frequent reference to 
    \citet[][hereafter \citetalias{2015ApJ...806...56O}]{2015ApJ...806...56O},
    who were the first to offer an explicit expression
    for the solution to the 2D~IVP.        
Noting that a continuous range of axial wavenumbers ($k$) is involved,
    Equation~(25) in \citetalias{2015ApJ...806...56O}
    expresses the solution as the summation of
    the contributions associated with an individual $k$,
    which in turn were written as the superposition
    of eigenmodes with individual 
    angular frequencies 
    ($\omega$, hereafter ``frequency'' for brevity).
A finite number of discrete $\omega$ pertaining 
    to proper eigenmodes (or ``trapped modes'' in physical terms) are 
    relevant only  when $k > k_{\rm cutoff, 1}$, whereby the periodicity
    is consistently $\lesssim 2\pi/(k_{\rm cutoff, 1} \vae) 
    \approx 2.6 \sqrt{1-\rhoe/\rhoi} (R/\vai) < 2.6 R/\vai$.    
Regardless of $k$, however,         
    a continuum of improper eigenmodes is always involved,
    the associated $\omega$ extending from $k\vae$ out to infinity.
The point is, only proper modes survive in          
    the sausage wavetrains sampled sufficiently far from 
    the exciter, the characteristic periodicities therefore being similar to
    the transverse \Alf\ time $R/\vai$.
If an individual $k$ is examined as happens for standing modes,
    then one finds by directly evolving the MHD equations 
    that $R/\vai$ consistently characterizes
        FSMs regardless of $k$
        \citep[e.g.,][]{2007SoPh..246..231T,2012ApJ...761..134N,2016SoPh..291..877G,2020ApJ...893...62L}.
While this result is much expected for $k > k_{\rm cutoff, 1}$,
	its physical understanding for the opposite situation is a bit involved
	given the likely contributions due to improper modes 
	with $\omega$ not far exceeding $k \vae$
	(\citetalias{2015ApJ...806...56O}, see also our Appendix~\ref{sec_app_step}).
The quick answer is that, 
    the interference of the improper modes tends
    to make their superposition favor a discrete
    set of periods $P_{\rm leaky}$ that pertain to the so-called ``leaky modes''
    \citep[and references therein]{2007PhPl...14e2101A},
    and $P_{\rm leaky}$ is well known to be either similar to or substantially shorter
    than $2\pi/(k_{\rm cutoff, 1} \vae)$
    \citep[e.g.,][]{1978SoPh...58..165M, 1986SoPh..103..277C,2007AstL...33..706K}.
The damping time of the discrete leaky modes ($\tau_{\rm leaky}$)
    is also known to offer a shortcut estimate 
    for the timescale characterizing the wave attenuation, the result
    being that $\tau_{\rm leaky}/P_{\rm leaky} \approx (\rhoi/\rhoe)/\pi^2$
    \citep[e.g.,][]{2007AstL...33..706K}.
Two primary reasons are now clear to account for
    the rarity of candidate coronal FSMs.        
First, $R/\vai$ typically evaluates to at most a couple of tens of seconds,
    thereby demanding a high instrumental cadence and ruling out
    the possibility for typical (extreme) ultraviolet instruments
    to resolve an FSM
     \citep[see][for exceptions]{2012ApJ...755..113S,2016ApJ...823L..16T}.
Second, there tends to be a stringent requirement on instrumental sensitivity as well.
For AR loops, that they are thin and tenuous means that FSMs tend to be detectable
    only as wavetrains, which are indeed compatible with a number of 
    high-cadence ground-based measurements in visible light during
    total eclipses \citep[e.g.,][]{2002MNRAS.336..747W,2003A&A...406..709K,2016SoPh..291..155S}.
For flare loops, that they are thick and dense means that FSMs
    have a better chance to be detected both as wavetrains and standing modes,
    provided once again that the instrumental cadence is sufficient
    \citep[see][and references therein]{2020SSRv..216..136L}.
This explains why candidate coronal FSMs reported so far
    have been primarily connected to radio measurements of short-period quasi-periodic pulsations 
    \citep[QPPs, see the recent reviews by][]{2018SSRv..214...45M,2021SSRv..217...66Z}. 

With the \citetalias{1983SoPh...88..179E} equilibrium apparently idealized, one may argue that cutoff wavenumbers
    are not inherent to coronal FSMs in reality. 
Indeed, there have been a considerable number of theoretical studies
    that extend \citetalias{1983SoPh...88..179E} by incorporating various aspects of reality
    \citep[see][and references therein]{2020SSRv..216..136L}.
Among these, we focus on the equilibria that differ from \citetalias{1983SoPh...88..179E} only by replacing the step
    density profile with a continuous one, the reason being that a generic guiding principle
    can be established to tell when cutoff wavenumbers exist 
    (\citealt[][hereafter \citetalias{2015ApJ...810...87L}]{2015ApJ...810...87L};
    also \citealt{2015ApJ...801...23L}).
Let $R$ now refer to some mean cylinder radius,
    and let the subscript ${\rm i}$ (${\rm e}$) refer to the equilibrium
    quantities at the cylinder axis (infinitely far from the cylinder).
The radial profile for the equilibrium density $\rho_0(r)$ can then be described
    in a generic form ${\rho_0}(r)= \rhoe+(\rhoi-\rhoe) f(r)$,
    where the function $f(r)$ evaluates to unity (zero) when $r=0$ ($r\to \infty$).
Restrict ourselves to the case where $f(r)$ is monotonical.
With Kneser's oscillation theorem, 
    \citetalias{2015ApJ...810...87L}
    were the first to point out that cutoff wavenumbers
    exist only when $r^2 f(r)$ does not diverge when $r$ approaches infinity.
This expectation was then verified numerically by \citet{2018ApJ...855...53L},
    one example being for the so-named ``outer $\mu$'' profile where
    $f(r)$ is identically unity for $r<R$ but of the form
    $(r/R)^{-\mu}$ otherwise. 
Figure~8 therein indicates that for the $m$-th radial harmonic ($m=1, 2, \cdots$),
    no cutoff wavenumber exists (or equivalently $k_{{\rm cutoff}, m} =0$)
    when $\mu <2$, whereas the combination  
    $(k_{{\rm cutoff}, m} R) \sqrt{\rhoi/\rhoe-1}$ increases monotonically
    from unity for $\mu=2$ to $j_{0, m}$ for a step profile ($\mu=\infty$).

Some important consequences arise for FSMs when cutoff wavenumbers are absent. 
Theoretically, the dispersive properties of FSMs in this situation
    are distinct from FSMs in \citetalias{1983SoPh...88..179E} in two aspects, one being that
    trapped modes are allowed regardless of the axial wavenumber $k$
    or the density contrast $\rhoi/\rhoe$, and the other being that
    FSMs tend to be weakly dispersive for small $k$ with axial phase speeds only marginally smaller than $\vae$
    (e.g., Figure~3 in \citetalias{2015ApJ...810...87L}, 
    and Figure~7 in \citealt{2017ApJ...836....1Y}).
With the former distinction evident, we note that trapped FSMs in \citetalias{1983SoPh...88..179E}
    are highly dispersive
    at least when $k$ is not far larger than a cutoff
    \citep[e.g.,][]{1983SoPh...88..179E,1983Natur.305..688R}.
Observationally, these two distinctions offer a richer possibility for interpreting
    oscillatory signals, to illustrate which we consider
    a spatially resolved QPP measured with the Nobeyama RadioHeliograph (NoRH)
    as reported by \citet{2013SoPh..284..559K}.
As detailed therein, multiple periodicities were simultaneously found, 
    with the associated 
    spatial distributions of the spectral power strongly indicating an axial fundamental
    together with its overtones in the involved flare loop.
Contrasting the observations with the canonical
	\citetalias{1983SoPh...88..179E} theory, the authors deduced that
    these standing modes belong to the kink family, and FSMs were ruled out because of
    their dispersive properties. 
However, adopting density profiles similar to the ``outer $\mu$'' one with $\mu <2$,
    both \citetalias{2015ApJ...810...87L} and \citet{2019MNRAS.488..660L} suggested that the observations
    may be interpreted as FSMs as well. 
Put to seismology, this interpretation returned values for the internal
    \Alf\ speed $\vai$ that may differ considerably from those returned
    with the interpretation in terms of kink modes, for which purpose we quote
    $\sim 1100~\kms$ from \citet[][section~6]{2019MNRAS.488..660L}  
    and $\sim 1750~\kms$ from \citet[][section~5.1]{2013SoPh..284..559K}.
Strictly speaking, a comparison between the two sets of $\vai$ 
    is not straightforward because
    \citet{2019MNRAS.488..660L} adopted pressureless MHD
    whereas a finite gas pressure is considered in \citet{2013SoPh..284..559K}
    \footnote{Given that flare loops tend to be dense and hot, one may question
        whether it is justifiable to adopt pressureless MHD.
        However, this issue is unlikely to be restrictive provided that the
        seismologically deduced $\vai$ is understood as
        the transverse fast speed \citep[see][]{2016ApJ...833..114C}.
        {However, caution needs to be exercised when one assesses 
        how FSMs are influenced by the curvature and lateral expansion
        of flare loops, both effects being observationally relevant
        but nonetheless not addressed here
        (for more discussions, see 
        e.g., \citealt{2009A&A...494.1119P} and \citealt{2016A&A...593A..52P} 
        as well as the references 
        both therein and in \citealt{2020SSRv..216..136L})}.}.
Our point is that the disappearance of cutoff wavenumbers for FSMs as a result of
    some straightforward departure of the equilibrium from \citetalias{1983SoPh...88..179E}
    can offer more physical interpretations for observations in the first place,
    and enable more seismological possibilities afterwards.

Some further consequence arises if we now focus on standing FSMs in AR loops.
For the ease of description, let us recall that
    we consistently work in pressureless MHD,
    and adopt the customary assumption that sees AR loops as straight, density-enhanced, field-aligned cylinders.
We further assume that the radial density distribution is of the ``outer $\mu$'' type, 
    which is reasonable but admittedly difficult to prove or disprove
    \citep[e.g.,][]{2003ApJ...598.1375A,2017A&A...605A..65G}.
In addition, we assume that lower coronal eruptions (LCEs), the primary exciter
    for the much-observed large-amplitude radial fundamental
        kink modes \citep{2015A&A...577A...4Z,2019ApJS..241...31N}, 
    can deposit a non-negligible amount of energy as axisymmetric perturbations 
    to AR loops as well. 
Note that this assumption is not that bold but has been implied in the interpretation 
    of rapidly propagating waves as sausage wavetrains
    \citep[see the review by][and references therein]{2008IAUS..247....3R}. 
For our purposes, it suffices to consider only axial fundamentals.
With dimensionless cutoff wavenumbers $\kcut R \ge 1/\sqrt{\rhoi/\rhoe-1}$ when $\mu\ge2$, 
    one finds that $\kcut R \ge 1/3$ 
    given the typical range of $[2, 10]$ quoted for
            the density contrast $\rhoi/\rhoe$ \citep{2004ApJ...600..458A}.
One further finds that             
    $kR = \pi R/L \lesssim \pi/15$ in view of the measurements 
        of widths and lengths for AR loops imaged in EUV
        \citep[][Figure~1]{2007ApJ...662L.119S}.  
Standing FSMs, at least axial fundamentals, are therefore unlikely to be 
    observable for two reasons.
One is that their periodicities will be $\sim R/\vai$ and therefore short,
    and the other is that they tend to experience rapid attenuation as well.       
Let us stress that these two signatures have not been explicitly shown for this
    particular ``outer $\mu$'' profile but 
    are expected with the studies on 1D IVPs addressing standing FSMs in the leaky regime
    for an \citetalias{1983SoPh...88..179E} equilibrium \citep[e.g.,][]{2007SoPh..246..231T,2012ApJ...761..134N}.
Now consider those AR loops with $\mu <2$.
Given the absence of cutoff wavenumbers, 
    the system is expected to settle to 
	a trapped mode or some combination of trapped modes, 
	the quality of the oscillatory signals
	being therefore sufficiently high.   
Likewise, the periodicites will be eventually 
    characterized by the longitudinal \Alf\ time 
    $L/\vae$, which can be readily resolved with, say, the majority
    of available UV/EUV instruments. 
In fact, both expectations have already been invoked in seismological applications,
    albeit in the context of flare loops (\citetalias{2015ApJ...810...87L}, \citealt{2019MNRAS.488..660L}).    
Supposing that AR loops with $\mu <2$ are not uncommon, one further deduces that
    a substantial fraction of kink oscillations will be mixed with standing FSMs
    when LCEs occur. 
As advocated by \citet{2015ApJ...812...22C} and \citet{2016SoPh..291..877G}, 
    the simultaneous observations of multiple modes of distinct nature
    will then considerably mitigate the non-uniqueness issue inherent
    to coronal seismology \citep[see][for dedicated remarks]{2019A&A...622A..44A}
    \footnote{
    Simultaneous observations of multiple modes are rare.
    Our point is that the more information
       one gathers from observations, 
       the better the to-deduce quantities can be constrained.
    Take the much-employed kink oscillations in AR loops.
    Their seismological applications significantly
       benefit from such additional information
       as the different damping characteristics
       in different stages of their temporal evolution
       \citep[e.g.,][]{2013A&A...551A..39H,
       2013A&A...555A..27R,2016A&A...585L...6P,2020ApJ...904..116G}.
    Likewise, additional information can be gleaned from the transverse distributions
       of the EUV emissions from AR loops
       (see \citealt{2018ApJ...860...31P} and also 
            \citealt{2017A&A...605A..65G}).
    On this aspect we note that a to-deduce parameter
       may still be constrained
       by such techniques as model averaging within the Bayesian framework
       even when observations do not favor one particular formulation
       out of many candidate formulations that involve this parameter
       \citep[see the review by][and references therein]{2018AdSpR..61..655A}.
    }.
However, standing FSMs have not been reported or even implicated in observations
    of oscillating AR loops to our knowledge.      
An obvious excuse is that observers have nearly exclusively adopted the \citetalias{1983SoPh...88..179E} framework
    and therefore dismissed the possibility that FSMs may possess periodicities
    $\gtrsim 2L/\vae$ altogether
	\footnote{
	The reason for FSMs to be excluded is more related to the fact that
	    the relevant periodicities were found in the transverse displacements of AR loops.
	See Appendix~\ref{sec_app_FD_scale} for more on this aspect.
	}.
Our point, however, is that this possibility is expected
    solely on the basis of EVP analyses on an open domain, and one has yet to demonstrate that
    FSMs with periodicities $\gtrsim 2L/\vae$ do exist as solutions to the
    pertinent IVP.

Focusing on sausage oscillations
    in AR loops with the ``outer $\mu$'' family of density profile, 
    we intend to address the question
    ``Does the absence of cutoff wavenumbers guarantee a temporal behavior
    that is distinct from the situation where cutoff wavenumbers are present?"
We decompose this question into two interconnected aspects.
One, how does the value of $\mu$ influence the timescale that
    characterizes the energy attenuation?
Two, does the transverse or longitudinal \Alf\ time characterize the periodicity 
    when a wave signal is sufficiently strong?       
This manuscript is structured as follows.
Section~\ref{sec_IVP_formu} formulates the IVP for a radially open system,
    which is then solved with a direct finite-difference (FD) approach
    in Section~\ref{sec_FD}.
While the answer to our question is already clear in the FD solutions,
    Section~\ref{sec_modal} moves on to re-solve the IVP by superposing eigen-solutions 
    to the relevant EVP on a closed domain. 
These modal solutions are presented for more than just cross-validation purposes.
Rather, they help quantify the specific contributions from individual frequencies.
By experimenting with various domain sizes, we will better connect the solutions
    to our IVP with the theoretical expectations from the analyses of EVPs on an open domain.
Section~\ref{sec_conc} summarizes this study, ending with some concluding remarks.

\section{Problem Formulation}
\label{sec_IVP_formu}
We adopt pressureless ideal MHD as our theoretical framework, 
   in which the primitive variables 
   are the mass density $\rho$, 
   velocity $\myvect{v}$,
   and magnetic field $\myvect{B}$. 
The equilibrium quantities are denoted with a subscript $0$, 
   and the equilibrium is taken to be static ($\myvect{v}_0 = 0$).
Working in a cylindrical coordinate system $(r, \theta, z)$,
   we take the equilibrium magnetic field to be uniform
   and directed in the $z$-direction 
   ($\myvect{B}_0 = B_0 \myvect{e}_z$). 
Seeing AR loops as density-enhanced cylinders with some mean radius $R$,
   we assume that the equilibrium density ($\rho_0$) depends only on $r$
   and decreases from $\rhoi$ at the cylinder axis ($r=0$)
   to $\rhoe$ infinitely far from the cylinder ($r\to\infty$). 
The \Alf\ speed is defined by $\va^2 = B_0^2/(\mu_0 \rho_0)$ with $\mu_0$ being
   the magnetic permeability of free space. 
From here onward, by ``internal'' (subscript ${\rm i}$)
   and ``external'' (subscript ${\rm e}$) we consistently 
   refer to the equilibrium quantities evaluated at $r=0$ and $r\to \infty$, respectively.
The internal (external) \Alf\ speed is therefore denoted by
   $\vai$ ($\vae$).

\subsection{Preliminary Formulation of the Initial Value Problem}
\label{sec_sub_probformu}

We now formulate the preliminary version of the 
    initial value problem (IVP) in a radially open system.
Let the subscript $1$ denote small-amplitude perturbations to the equilibrium.
Specializing to sausage perturbations ($\partial/\partial \theta = 0$),
    the linearized, pressureless, ideal MHD equations read
\begin{eqnarray}
   \rho_0 \dispfrac{\partial v_{1r}}{\partial t}
&=& 
   \dispfrac{B_0}{\mu_0}
   \left( 
          \dispfrac{\partial B_{1r}}{\partial z}
        - \dispfrac{\partial B_{1z}}{\partial r}
   \right), 
					\label{eq_linMHD_momen_comp_r} 
\\
   \dispfrac{\partial B_{1r}}{\partial t}
&=&  
   B_0 \dispfrac{\partial v_{1r}}{\partial z}, 
				    \label{eq_linMHD_Farad_comp_r} 
\\ 
   \dispfrac{\partial B_{1z}}{\partial t}
&=&  
   -B_0 
    \dispfrac{1}{r}
	\dispfrac{\partial }{\partial r} \left( r v_{1r} \right)~. 
	      			\label{eq_linMHD_Farad_comp_z} 
\end{eqnarray}     
With coronal cylinders bounded by the planes $z=0$ and $z=L$ in mind,
   the following ansatz
\begin{equation}
\label{eq_Fourier_ansatz}
\begin{split}
  v_{1r} (r, z; t) 
& = \hat{v}  (r; t) \sin(kz), 
\\
  B_{1r} (r, z; t) 
&= \hat{B}_r(r; t) \cos(kz), 
\\
  B_{1z} (r, z; t) 
&= \hat{B}_z(r; t) \sin(kz), 
\end{split}
\end{equation}     
   is appropriate for axial standing modes,
   with $k = n \pi/L$ being the quantized 
   axial wavenumber ($n=1, 2, \cdots$).
Equations~\eqref{eq_linMHD_momen_comp_r} to \eqref{eq_linMHD_Farad_comp_z} 
   then become
\begin{eqnarray}
    \rho_0 \dispfrac{\partial \hat{v}}{\partial t}
&=& 
   -\dispfrac{B_0}{\mu_0}
    \left( 
          k \hat{B}_r 
        + \dispfrac{\partial \hat{B}_{z}}{\partial r}
    \right), 
			\label{eq_linMHD_momen_comp_r_Fourier} 
\\
   \dispfrac{\partial \hat{B}_{r}}{\partial t}
&=&  
   k B_0 \hat{v}, 
	      	\label{eq_linMHD_Farad_comp_r_Fourier} 
\\
   \dispfrac{\partial \hat{B}_{z}}{\partial t}
&=&  
   -B_0 
	\dispfrac{1}{r}
	\dispfrac{\partial }{\partial r} \left( r \hat{v} \right)~. 
	      \label{eq_linMHD_Farad_comp_z_Fourier} 
\end{eqnarray} 
Without loss of generality, the initial conditions (ICs) are specified as
\begin{eqnarray}
&&  \hat{v}(r, t=0) = u(r), \label{eq_IC_v}\\
&&  \hat{B}_r (r, t=0)
  = \hat{B}_z (r, t=0)
  = 0. \label{eq_IC_B} 
\end{eqnarray}
The boundary condition (BC) at    
    the cylinder axis ($r=0$) reads
\begin{eqnarray}
\hat{v} = \hat{B}_r = \partial\hat{B}_z/\partial r = 0,
	\label{eq_BCs}  
\end{eqnarray}
    whereas the BC at $r\to\infty$ is irrelevant. 

It proves necessary to examine the energetics associated with
    the IVP as well.
Let $V$ refer to a volume $V$
   bounded laterally by a cylindrical surface with radius $r$
   and horizontally by the planes $z = 0$ and $z=L$.
One then finds from     
   Equations~\eqref{eq_linMHD_momen_comp_r_Fourier} to 
   			 \eqref{eq_linMHD_Farad_comp_z_Fourier}
   that 
\begin{eqnarray}
     E_{\rm tot}(r, t) - E_{\rm tot}(r, t=0) 
  = -F(r, t),
  	\label{eq_ConsInt_Etot}
\end{eqnarray}           
   where 
\begin{eqnarray}
     E_{\rm tot}(r, t)    
&=&  
     \pi L \int_{0}^{r} (r' dr')
       \left\{
          \dispfrac{1}{2}\rho_0 (r') \hat{v}^2(r', t)
         +\dispfrac{1}{2\mu_0}
                   \left[ \hat{B}_r^2(r', t)
                        + \hat{B}_z^2(r', t)
                   \right] 
       \right\}~, 
     \label{eq_def_ener_Etot} 
\\
	F(r, t) 
&=& 
    \pi L\int_{0}^{t} dt'
	\left[r \hat{p}_{\rm tot}(r, t') \hat{v} (r, t')\right]~.   
	\label{eq_def_ener_F}           
\end{eqnarray}  
Here a common factor $\pi L$ is retained to ensure that
    $E_{\rm tot}(r, t)$ represents the instantaneous total energy in $V$,
    while $F(r, t)$ represents the cumulative energy loss from $V$.
Furthermore, $\hat{p}_{\rm tot} = B_0 \hat{B}_z/\mu_0$ is
    connected to the Eulerian perturbation of total pressure. 
Evidently, the terms in the square parentheses in 
    Equation~\eqref{eq_def_ener_F} stem from the radial component
    of the Poynting vector.    

\subsection{Reformulation of the IVP and Parameter Specification}
For mathematical convenience,  
	Equations~\eqref{eq_linMHD_momen_comp_r_Fourier} to 
	\eqref{eq_BCs} are reformulated  
    to the following form.
\begin{IVP} 
\label{ivp_mu_open}
Solutions are sought for the following equation 
   \begin{eqnarray}
   \displaystyle 
      \frac{\partial^2 \hat{v}}{\partial t^2}
   &=&
      \va^2(r)
      \left(
          \frac{\partial^2 \hat{v}}{\partial r^2}
        + \frac{1}{r}
          \frac{\partial \hat{v}}{\partial r}
        - \frac{\hat{v}}{r^2}  
        - k^2 \hat{v}    
      \right),  \label{eq_v2nd_final}    
   \end{eqnarray}
   subjected to the ICs
\begin{eqnarray}
\hat{v}(r, t=0) = u(r), \quad  
\frac{\partial \hat{v}}{\partial t} (r, t=0) = 0,
\label{eq_v2nd_IC}
\end{eqnarray}
   togeter with the BC
\begin{eqnarray}
\hat{v}(r=0, t) = 0, \label{eq_FD_BC}
\end{eqnarray}   
   on a domain spanning from $r=0$ to $\infty$.
\end{IVP}
Necessary for energetics considerations,
   $\hat{B}_r(r, t)$ and $\hat{B}_z(r, t)$ can be found 
   with $\hat{v}(r, t)$ by integrating 
   Equations~\eqref{eq_linMHD_Farad_comp_r_Fourier} and 
     \eqref{eq_linMHD_Farad_comp_z_Fourier}
   from the initial state~\eqref{eq_IC_B}. 
        
We proceed to make IVP~\ref{ivp_mu_open} more specific.
The equilibrium density distribution is chosen to be
    the ``outer $\mu$'' profile in \citet{2017ApJ...836....1Y}, namely
\begin{equation}
\label{eq_rho_prof_outermu}
\begin{split} 
& {\rho_0}(r)
   = \rhoe+(\rhoi-\rhoe) f(r), 
\\
&  f(r) 
   =
    \left\{
       \begin{array}{ll}
          1,     						& ~~0 \le r \le R, 			\\[0.1cm]
          \left(r/R\right)^{-\mu}, 	    & ~~r \ge R.
       \end{array} 
    \right.
\end{split}   
\end{equation}
Here $\mu \ge 1$ measures the steepness of $\rho_0(r)$ outside the cylinder.
We focus on axial fundamentals ($k = \pi/L$). 
In addition, we specify the initial perturbation
    in Equation~\eqref{eq_v2nd_IC} as
\begin{eqnarray}
\frac{u(r)}{\vai} =  
    \left\{
       \begin{array}{ll}
          \sin^3(\pi r/\Lambda),   & \quad 0 \le r \le \Lambda, 			\\[0.1cm]
          0, 	    			   & \quad r \ge \Lambda,
       \end{array} 
    \right.
 \label{eq_u}
\end{eqnarray}    
    which is localized within $r=\Lambda$
    and prescribed to be sufficiently smooth with a magnitude arbitrarily
    set to be the internal \Alf\ speed ($\vai$).

\begin{figure}
\centering
 \includegraphics[width=.9\columnwidth]{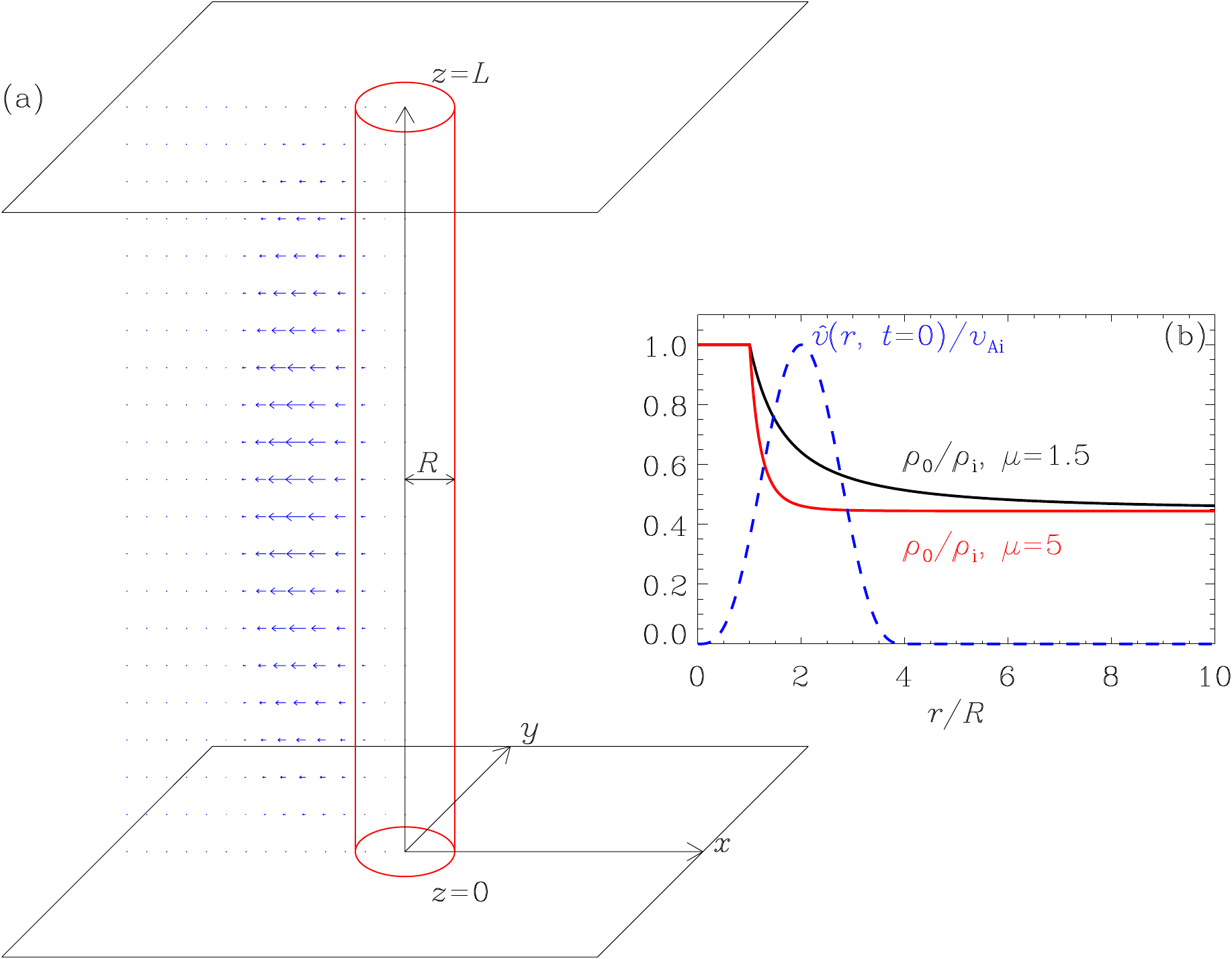}
 \caption{
(a) Illustration of the equilibrium configuration,
     together with the initial velocity field in an arbitrary
     plane through the cylinder axis (the blue arrows).
The $z$-dependence of the initial perturbation leads to
    axial fundamentals.     
(b) Radial profiles of the initial perturbation
    ($\hat{v}$, the blue dashed curve) 
    and the equilibrium density ($\rho_0$, the solid curves),
    both involved in IVP~\ref{ivp_mu_open}.
The density contrast $\rhoi/\rhoe$ is chosen to be $2.25$.    
Two steepness parameters are examined, namely
    $\mu = 1.5$ (the black curve)
    and $\mu = 5$ (red).
 }
 \label{fig_EQprofile}
\end{figure}

The solution to IVP~\ref{ivp_mu_open} 
    is fully determined by the dimensionless paramters
        $[\rhoi/\rhoe, \mu; L/R; \Lambda/R]$, among which
    we see $\mu$ as the primary adjustable one.     
The density contrast and loop length-to-radius ratio are
    fixed at $[\rhoi/\rhoe, L/R] = [2.25, 15]$,
    both close to the lower end of but nonetheless within
    the accepted range for AR loops
    \citep[e.g.,][]{2004ApJ...600..458A,2007ApJ...662L.119S}.
We take $\Lambda = 4~R$ unless otherwise specified.        
Figure~\ref{fig_EQprofile}a illustrates our equilibrium, 
    and the blue arrows represent the initial velocity field in
    any cut through the cylinder axis as appropriate for an axial fundamental. 
Specializing to IVP~\ref{ivp_mu_open}, Figure~\ref{fig_EQprofile}b shows
    the radial profiles for $u(r) = \hat{v}(r, t=0)$ (the blue dashed curve)
    and for the equilibrium density ($\rho_0$, the solid curves).  
Two values are adopted for $\mu$, one being $1.5$
    (the black curve) and the other being $5$ (red).      
As already stressed, FSMs do not suffer from
    cutoff wavenumbers $\kcut$ for $\mu <2$.
When $\mu \ge 2$, $\kcut R$ always exceeds 
    $1/\sqrt{\rhoi/\rhoe-1}$ and therefore $\gtrsim 0.89$
    with the chosen $\rhoi/\rhoe$, 
    making trapped modes irrelevant for the chosen $kR = \pi/15 \approx 0.21$.

\section{Finite-Difference Solutions}
\label{sec_FD}      
We choose to solve IVP~\ref{ivp_mu_open} with two independent methods,
    one being a finite-difference (FD) approach,
    and the other being a modal approach involving eigenmodes
    for the EVPs on either a closed or an open domain.
In practice, the FD approach turns out to be orders-of-magnitude
    less time-consuming, 
    and is therefore more suitable for parametric studies. 
    
\subsection{Method} 
The development of our FD code starts with constructing a system of code units, 
    the details of which
    are irrelevant because we will consistently present our results as dimensional quantities.    
Equation~\eqref{eq_v2nd_final}
    is discretized on a uniform grid with spacing $\Delta r = 0.01~R$
    over a domain of $[0, r_{\rm M}]$.
All spatial derivatives are approximated by
    centered differences, yielding second-order accuracy in space.
The time-marching is handled in a leap-frog manner with a uniform timestep
    $\Delta t$. 
A ghost timestep at $t=-\Delta t$ is introduced
    to account for the IC for $\partial \hat{v}/\partial t$, 
    ensuring that the scheme is second-order
    accurate in time as well. 
The timestep is specified as $\Delta t = c \Delta r/\vae$, 
    where the Courant number $c$ is chosen to be $\sim 0.4$ to ensure
    numerical stability. 
Grid convergence tests are conducted to ensure that varying the
    grid spacing or the Courant number does not influence our results.
More importantly, we make sure that the location of
    the outer boundary $r_{\rm M}$ does not affect
    our FD solutions either.

\subsection{Numerical Results}  
\begin{figure}
\centering
 \includegraphics[width=.8\columnwidth]{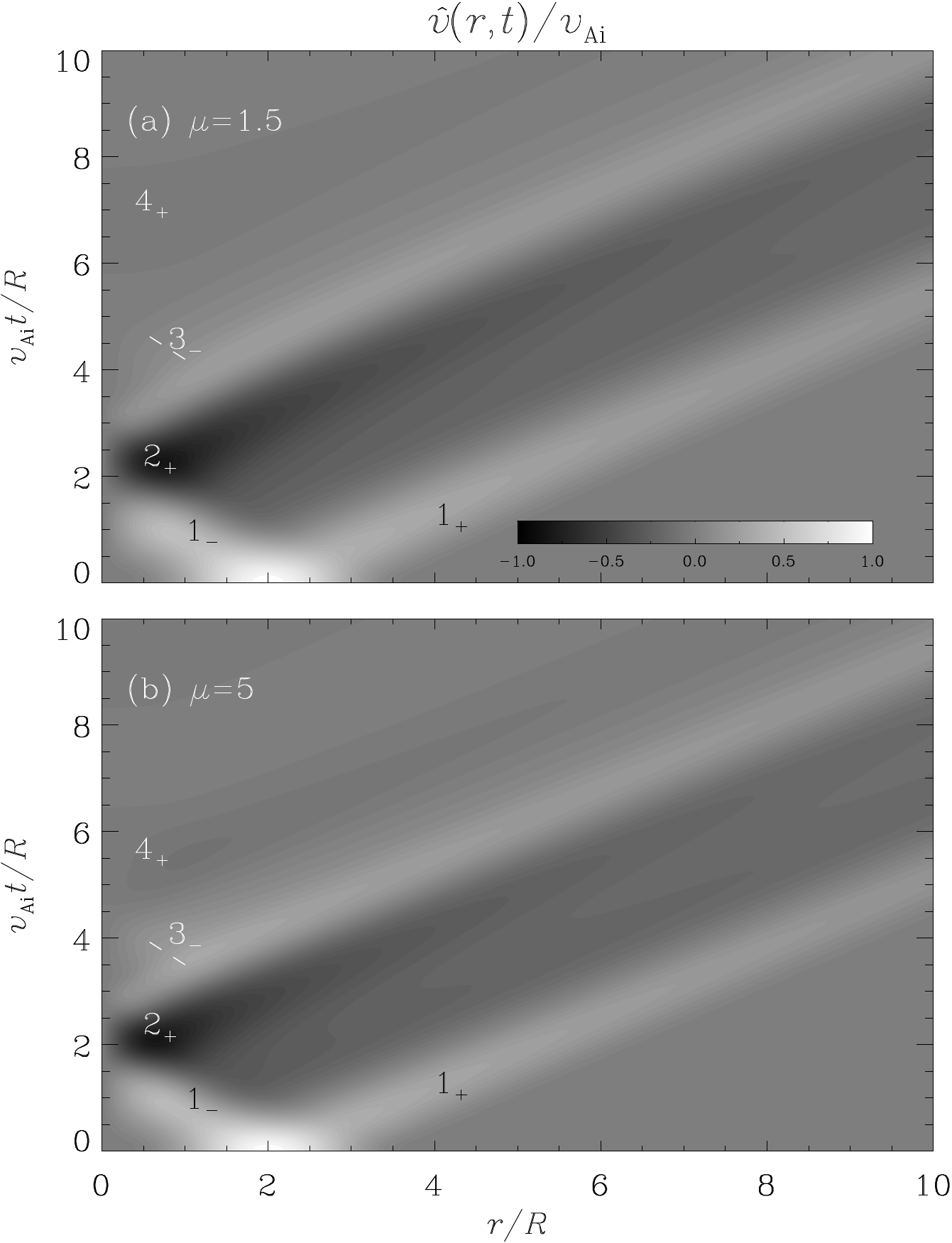}
 \caption{
 Finite-difference (FD) solutions to IVP~\ref{ivp_mu_open}
     for equilibrium density profiles with
     the steepness parameter being (a) $\mu=1.5$
     and (b) $\mu=5$.
 Plotted are the distributions of the radial speed $\hat{v}$ 
 	 in the $r-t$ plane. 
 Some prominent wavefronts are singled out as labeled, with the plus (minus)
     sign representing outward (inward) propagation.
 Both computations pertain to the combination 
     $[\rhoi/\rhoe, kR, \Lambda/R] = [2.25, \pi/15, 4]$.    
 }
 \label{fig_vsnapshots}
\end{figure}

Figure~\ref{fig_vsnapshots} presents our FD solutions by showing
    the distribution of the radial speed $\hat{v}$
    in the $r-t$ plane.  
We contrast two cases with the steepness parameter being
    (a) $\mu=1.5$ and (b) $\mu = 5$, respectively.
{One sees from Figure~\ref{fig_vsnapshots}} that 
    the temporal evolution in both cases is characterized 
    by some dispersive propagation of the axisymmetric disturbance.
To proceed, we note that the disturbances belong to the fast family,
    and the local fast speed is equivalent to the local \Alf\ speed 
    in pressureless MHD. 
For the ease of description, the most prominent wavefronts are singled out
    and labeled such that
    the $+$ ($-$) sign pertains to outward (inward) propagation. 
When necessary, the same label will be used to denote the associated wake 
    as well.
The following features can then be readily told in both cases,
    to describe which it suffices to consider only Figure~\ref{fig_vsnapshots}a.
Firstly, the initial perturbation splits into two wavefronts 
    manifested as the two bright stripes labeled $1_{+}$ and $1_{-}$.
While not that evident, wavefront $1_{-}$ is actually accompanied by
    a wake appearing as a narrow dark stripe.      
At least a substantial fraction of both wavefront $1_{-}$ and its wake
    then make it into the cylinder ($r<R$),
    as evidenced by the change of slope of the stripe. 
Here by ``a substantial fraction'' we mean that some reflection
    is expected but difficult to identify.
Secondly, once reaching the cylinder axis ($r=0$), wavefront $1_{-}$ is reflected to
    form wavefront $2_{+}$ which then propagates outward as a 
    dark stripe.
The change from a bright to a dark stripe for essentially the same wavefront
    is simply because $\hat{v}$ necessarily reverses sign 
    at the cylinder axis, which acts as a rigid wall in the present context
    \citep[see also][hereafter \citetalias{1996ApJ...472..398B}]{1996ApJ...472..398B}.
In this sense, the bright stripe following wavefront $2_{+}$, namely wake $2_{+}$,
     is actually the reflected wake $1_{-}$.
Thirdly, the partial reflection of wake $2_{+}$ around the cylinder boundary ($r=R$)
    then leads to wavefront $3_{-}$, part of which is guided by
    the dashed curve.
Wavefront $4_{+}$ then result from the reflection of wavefront $3_{-}$ at 
    the cylinder axis. 
Having described these common features, we note that some differences
    nonetheless exist between the two cases.
For instance, wavefront $3_{-}$ is easier to tell for $\mu=5$
    than for $\mu=1.5$.
This is understandable because the case with $\mu=5$ corresponds to
    a steeper $\va$ profile around the cylinder boundary
    and hence a stronger partial reflection there.

\begin{figure}
\centering
 \includegraphics[width=.7\columnwidth]{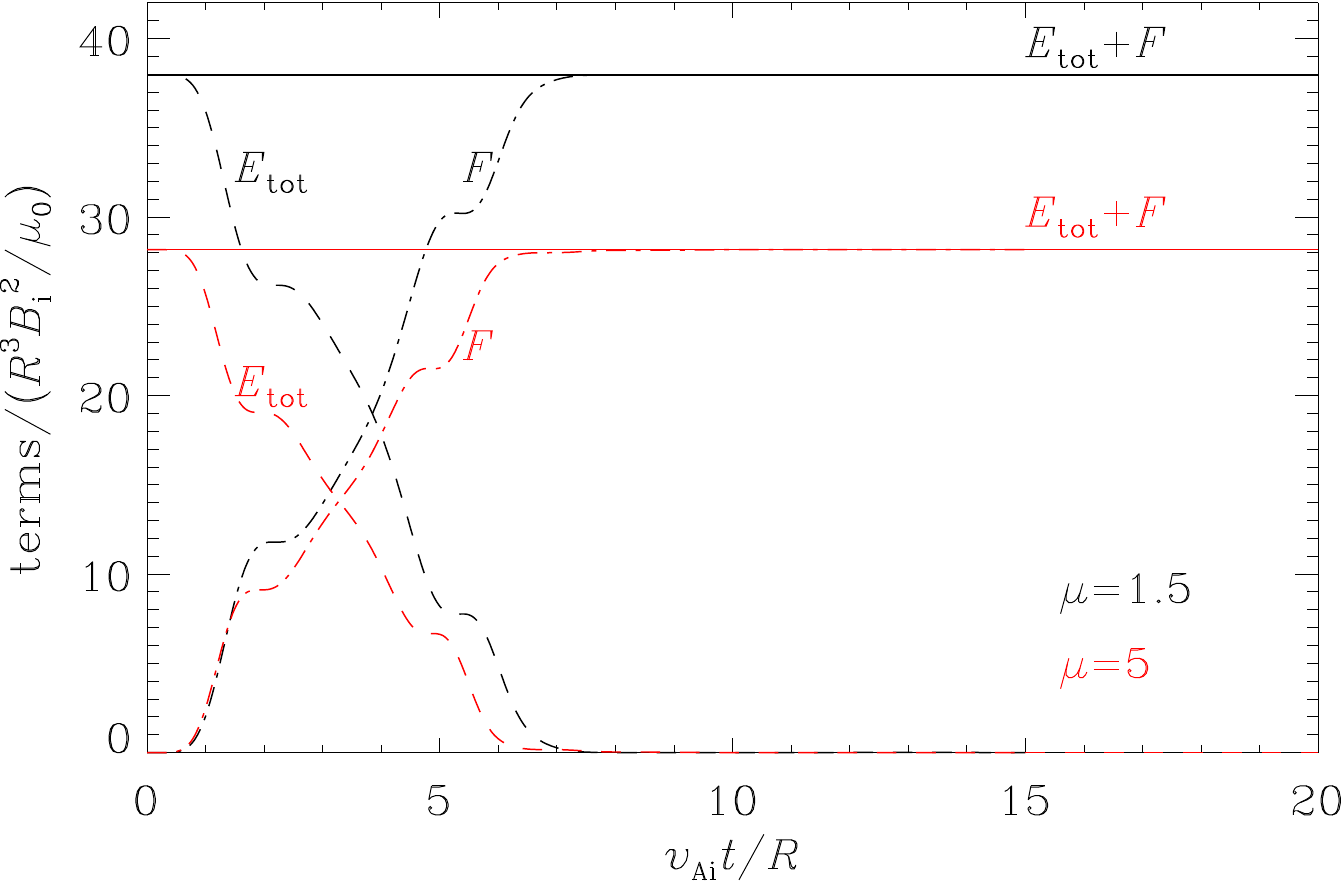}
 \caption{
 Temporal variations of some terms characterizing the 
     energetics of axisymmetric perturbations in the cylindrical volume ($V$)
     where the initial perturbation is applied. 
 The dashed curves labeled $E_{\rm tot}$ represent the total wave energy in $V$,
     while the dash-dotted curves represent the cumulative energy loss from $V$
           and are labeled $F$ (see Equations~\eqref{eq_def_ener_Etot} and
           \eqref{eq_def_ener_F} for definitions).       
 Their sum is plotted by the solid curves. 
 The energetics terms are evaluated with the finite-difference
     solutions to IVP~\ref{ivp_mu_open} 
     for two steepness parameters, 
     one being $\mu=1.5$ (the black curves)
     and the other being $\mu=5$ (red).
 Both computations pertain to the combination 
     $[\rhoi/\rhoe, kR, \Lambda/R] = [2.25, \pi/15, 4]$.    
 }
 \label{fig_enerFD}
\end{figure}

The slight differences notwithstanding, in both cases one expects a continuous decrease
    for the wave energy in the cylindrical volume $V$ bounded by $r = \Lambda$,
    namely where the initial perturbation is applied. 
This expectation is indeed true
    \footnote{
    In addition to $\mu$, the value of $\Lambda$ is also varied
        in the parametric survey to be presented shortly. 
    When $\Lambda \gtrsim 6~R$, we can discern some very brief time intervals
        with widths $\lesssim 0.2 R/\vai$ during which
        the total energy $E_{\rm tot}$ shows an extremely weak increase.
    However, these intervals are not important for our purposes because they
        appear exclusively after $E_{\rm tot}$ has 
        already decreased by a factor of $\gtrsim 50$.}, 
    to demonstrate which    
    we display the temporal variations of 
    the total energy in $V$ ($E_{\rm tot}$, the dashed curves)
    and the cumulative energy loss from $V$ ($F$, dash-dotted) 
    in Figure~\ref{fig_enerFD}. 
The sum $E_{\rm tot}+F$ is further given by the solid curves.
We discriminate between the cases with $\mu = 1.5$ and $\mu = 5$ 
    by different colors. 
From the solid curves one sees that energy conservation is maintained 
    remarkably well, to quantify which we quote an accuracy of better than
    $0.05\%$ for all FD computations. 
Examining the dashed curves, one sees that $E_{\rm tot}$ 
    shows a couple of plateaus with the behavior of $E_{\rm tot}$ for $\mu=5$
    around $t \sim 5~R/\vai$ being an example. 
In view of Equations~\eqref{eq_ConsInt_Etot} and \eqref{eq_def_ener_F},
    these plateaus appear simply because the radial component of the Poynting flux
    tends to vanish therein.
More importantly, $E_{\rm tot}$ rapidly decreases with time,
    with the two most prominent intervals readily accounted
    for by the passage of wavefronts $1_{+}$ and $2_{+}$ (see Figure~\ref{fig_vsnapshots}).
Virtually no energy is left in $V$ when $t \gtrsim 7~R/\vai$,    
    which is true for $\mu = 5$ and $\mu=1.5$ alike.
    
The rapid attenuation of wave energy can be told, in a more straightforward way,
    by directly showing the temporal evolution of the
    radial speed itself $\hat{v}$. 
This is done in Figure~\ref{fig_vtdep_FD_vs_modal} where we plot $\hat{v}$
    at $r=R$ and use different colors
    to discriminate different values of $\mu$. 
We note that the FD solutions are shown by the solid curves, labeled ``FD open" to
    reflect that they are found on an open domain.
The asterisks, labeled ``modal closed'', represent the solutions
    found by superposing the eigenfunctions for the EVP on a closed domain.
The details of the modal solutions are not important for now.
What matters is that they agree with the FD ones exactly, thereby suggesting the reliability
    of both sets of solutions.
Consider now only the solid curves. 
With the aid of Figure~\ref{fig_vsnapshots}, 
    one readily identifies the first three extrema with
    wavefronts $1_{-}$ and $2_{+}$ as well as wake $2_{+}$. 
One further sees one more extremum (two more extrema) in the black (red) curve,
    the corresponding wavefronts/wakes also identifiable in Figure~\ref{fig_vsnapshots}.
When discernible, any extremum in the black curve appears later than
    its counterpart in the red curve.
The explanation for this behavior is actually straightforward because
    the extrema in $\hat{v}(R, t)$ ultimately derive from
    wavefront $1_{-}$.
One sees from Figure~\ref{fig_EQprofile} that     
    the local fast speed $\va(r)$ at any $r > R$
    is larger for $\mu=5$ than for $\mu=1.5$.
It therefore takes more time for wavefront $1_{-}$ to enter the cylinder ($r\le R$)
    in the case with $\mu=1.5$, thereby making the relevant 
    extrema in $\hat{v}(R, t)$ appear later. 
With this understanding, the spacing between two consecutive prominent extrema,
    or equivalently the periodicity when the signal is strong, 
    is intimately connected to the passage of the relevant
    wavefronts/wakes before they appear in $\hat{v}(R, t)$.
The end result is that, the periodicity is expected to depend on the details of 
    both the equilibrium density profile
    and the initial perturbation.
In our setup, this translates into the dependence on $\mu$ and $\Lambda$.     
     
Figure~\ref{fig_survey_mu_Lambda} quantifies the dependence of the wave behavior
    on the steepness parameter ($\mu$) for a number of values of
    the spatial extent of the initial perturbation ($\Lambda$) as labeled.
Two quantities are examined, one being the time that it takes
    for $E_{\rm tot} (\Lambda, t)$ to drop from the initial value
    by a factor of ${\rm e}^{4} \approx 55$
    ($\tauener$, Figure~\ref{fig_survey_mu_Lambda}a)
    and the other being the temporal spacing between the first two extrema
    in the $\hat{v} (R, t)$ profile 
    ($\Delta_1$, Figure~\ref{fig_survey_mu_Lambda}b). 
Let us examine Figure~\ref{fig_survey_mu_Lambda}a first,
	and start by noting that 
	we deliberately choose a rather large factor (${\rm e}^{4}$)    
    to determine $\tauener$.
We note further that the initial perturbation peaks at $r = \Lambda/2$
    (see Equation~\eqref{eq_u}).    
Now two prominent features are evident. 
Firstly, $\tauener$ at any given $\mu$ increases with $\Lambda$.
Secondly, the $\mu$-dependence of $\tauener$
	tends to be weak when $\Lambda \lesssim 2~R$.
Let $V$ still denote the cylindrical volume laterally bounded by $r=\Lambda$.
It turns out that the departure of wavefronts $1_{+}$ and $2_{+}$ from $V$
    is the primary reason for $E_{\rm tot} (\Lambda, t)$ 
    to decrease to the designated threshold
    (see Figure~\ref{fig_vsnapshots}). 
In particular, $\tauener$ tends not to be
    much longer than the time $t_{2+}$ at which
    wavefront $2_{+}$ arrives at $r=\Lambda$.
In turn, this transit time $t_{2+}$ comprises two components, 
\begin{eqnarray}
t_{2+} = t_{1-} (\Lambda/2 \to 0) + t_{2+} (0 \to \Lambda),
\label{eq_tmp1}
\end{eqnarray}
    where the symbols on the right-hand side (RHS) are such
    that $t_{1-} (\Lambda/2 \to 0)$
    represents the time that wavefront $1_{-}$ spends when 
    traveling from $r=\Lambda/2$ to $r=0$. 
For a given $\mu$, both terms on the RHS
    increase with $\Lambda$, meaning that
    $t_{2+}$ and hence $\tauener$ increase monotonically with $\Lambda$.
Now move on to the $\mu$-dependence for a given $\Lambda$.     
When $\Lambda \lesssim 2~R$, the cylinder exterior
   ($r >R$) is relevant for determining $t_{2+}$ only by 
   being partially involved in $t_{2+} (0 \to \Lambda)$.
When $\Lambda > 2~R$, however, it is involved
   in both terms on the RHS of Equation~\eqref{eq_tmp1}.
The end result is that, $t_{2+}$ and hence $\tauener$ are  
    insensitive to $\mu$ when $\Lambda \lesssim 2~R$ but tend to decrease with $\mu$ when the opposite is true, which is understandable given that
    the local fast speed $\va(r)$ at any $r>R$ increases with $\mu$.

Now move on to Figure~\ref{fig_survey_mu_Lambda}b. 
One sees that $\Delta_1$ possesses a considerably more complicated
    behavior, 
    by which we mean the features difficult to understand
    with the simple kinematic  considerations that were applied to Figure~\ref{fig_survey_mu_Lambda}a.
Take the cases where $\Lambda = 4~R$ and $\Lambda = 8~R$.
In both cases, the first and second extrema in the $\hat{v}(R, t)$
    profile correspond to wavefronts $1_{-}$ and $2_{+}$, respectively. 
In kinematic terms, the temporal spacing between the two
    then comprises $t_{1-}(R \to 0)$ and $t_{2+}(0 \to R)$, 
    neither of which is supposed to involve $\Lambda$ or $\mu$.
Consequently, the blue and maroon curves are expected to 
    overlap, an expectation evidently at variance with the
    numerical results. 
One therefore deduces that $\Delta_1$ embeds some subtleties 
    that the kinematic arguments cannot address,
    which will become evident in the modal solutions
    to be presented shortly.
The quick answer is the common sense that $\mu$ is relevant for determining
    the eigenstructures, while $\Lambda$ determines how the energy in the initial
    perturbation is distributed among the eigenmodes
    (see Equation~\eqref{eq_modal_formalSol}). 
Important for now is that Figure~\ref{fig_survey_mu_Lambda}
    has already answered the questions we laid out
    in the Introduction.
Firstly, Figure~\ref{fig_survey_mu_Lambda}a indicates that the energy imparted
    by the initial perturbation is attenuated within a characteristic timescale
	$\tauener \sim \bigO(\Lambda/\vai)$.
If $\Lambda$ is not far different from $R$, then axisymmetric 
    perturbations will rapidly become too weak to detect.
Secondly, even if some instrument happens to capture
    a perturbation immediately after its excitation, 
    Figure~\ref{fig_survey_mu_Lambda}b indicates that 
    $\Delta_1$ is consistently $\sim \bigO(R/\vai)$, thereby placing
    rather stringent demands on the instrumental cadence. 
Thirdly, for any examined $\Lambda$,  
    no abrupt change is seen in the behavior of $\tauener$ or $\Delta_1$ 
    when $\mu$ crosses the nominally critical value of two. 
Put another way, trapped modes are not discernible even though
    they suddenly appear when $\mu$ drops below two in EVPs
    on an open domain. 
We address why in what follows.

\section{Modal Solutions}
\label{sec_modal}

\subsection{Method}
Our modal approach starts with specifying the following EVP on a
    closed domain. 
\begin{EVP} 
\label{evp_mu_closed}
Nontrivial solutions are sought for the following equation 
\begin{equation}
  - \omega^2 \breve{v} 
=  \va^2(r)
   \left( 
		  \dispfrac{d^2}{dr^2}\breve{v} 
		+ \dispfrac{d}{r dr  }\breve{v}
		- \dispfrac{\breve{v}}{r^2} 
		- k^2 \breve{v}
   \right), 
\label{eq_modal_EVP_govE}
\end{equation}
   defined on a domain of $[0, d]$ and subjected to the BCs
\begin{equation}
\breve{v} (r=0) = \breve{v} (r=d) = 0.
\label{eq_modal_EVP_BC}
\end{equation} 
\end{EVP}
Equation~\eqref{eq_modal_EVP_govE} is found by replacing $\hat{v}$ with
    ${\rm Re}[\breve{v}(r) \exp(-i\omega t)]$ in
    Equation~\eqref{eq_v2nd_final}.

EVP~\ref{evp_mu_closed} is known to 
    possess the following Sturm-Liouville properties
    (see \citetalias{1996ApJ...472..398B} for details, even though
    a step profile is examined therein).
First of all, the eigenvalues ($\omega^2$) are positive and form
    an infinite, discrete, monotonically increasing sequence $\{\omega_{l}\}$ with respect
    to the mode number $l = 1, 2, \cdots$. 
The associated eigenfunction $\breve{v}_l(r)$
	can be made and will be seen as real-valued.
It then follows that $\breve{v}_l(r)$ possesses $l-1$ nodes inside the domain.
In addition, the set $\{\breve{v}_l(r)\}$ is complete and satisfies the orthogonality condition
\begin{eqnarray}
\int_{0}^{d}
   \breve{v}_l (r) \breve{v}_m (r) \rho_0(r) rdr 
= 0,
\end{eqnarray}    
   provided $l \ne m$.
Eventually, the solution to IVP~\ref{ivp_mu_open} can be written as
\begin{equation}
\label{eq_modal_formalSol}
\begin{split}
&  \vhatd (r, t)
  = \sum\limits_{l = 1}^{\infty} 
        c_l  \breve{v}_l(r) \cos(\omega_l t), 
\\
& 0 \le r \le d, \quad         
  t \le \int_{\Lambda}^{d} \dispfrac{dr}{\va(r)},
\end{split}
\end{equation}
    where the coefficient $c_l$ measures the contribution
    from the $l$-th mode,
\begin{eqnarray}
  c_l 
= 
  \frac{\displaystyle \int_{0}^{d}
  		   u(r) \breve{v}_l(r) \rho_0(r) rdr}
       {\displaystyle \int_{0}^{d}
           \breve{v}^2_l (r) \rho_0(r) rdr}.
  \label{eq_modal_coef}
\end{eqnarray}
The superscript $(d)$ in Equation~\eqref{eq_modal_formalSol} 
   is meant to indicate that
   the modal structure depends on the domain size $d$.
Here and hereafter, by ``modal structure" we further mean 
   the $l$-dependence of $\omgl$.
Expressed formally, $\omgl$ can be written as
\begin{equation}
\label{eq_EVP_omgl_formal}
   \dispfrac{\omega_{l} R}{\vai} 
=  {\mathcal{F}}_l(\rhoi/\rhoe, \mu; k R; d/R).
\end{equation} 
However, we stress that the modal solution $\vhatd(r, t)$ itself
   does not depend on $d$ in the timeframe of validity explicitly
   given in Equation~\eqref{eq_modal_formalSol}.
Physically, this timeframe of validity represents simply the interval when
   the outermost edge of the perturbation is within the domain of EVP~\ref{evp_mu_closed}.
      
We now describe some details involved in the evaluation of 
   Equation~\eqref{eq_modal_formalSol}.
To start, $[\rhoi/\rhoe, kR, \Lambda/R]$ is fixed at $[2.25, \pi/15, 4]$
   for all modal solutions.
In the main text, we contrast only two steepness parameters ($\mu=1.5$ and $\mu=5$)
   but experiment with a substantial number of dimensionless domain sizes ($d/R$).
The step profile ($\mu=\infty$) is also of interest, 
   the discussions nonetheless collected in Appendix~\ref{sec_app_step}. 
Regardless, we consistently formulate and solve EVP~\ref{evp_mu_closed}
   with the general-purpose finite-element
   code PDE2D \citep{1988Sewell_PDE2D},
   which was first introduced into the solar context
   by \citet{2006ApJ...642..533T}.
A uniform grid is adopted if $d/R \le 150$, 
   otherwise we employ a grid whereby
   the spacing is uniform for $r \le 5~R$ but increases by a constant factor
   afterwards. 
We make sure that different grid setups yield consistent solutions
   to EVP~\ref{evp_mu_closed}. 
Likewise, we make sure that the number of modes ($l_{\rm max}$) 
   incorporated in the summation in Equation~\eqref{eq_modal_formalSol}
   is sufficiently large, meaning that
   increasing $l_{\rm max}$ does not influence the modal solutions
   to be analyzed.     

Some insights into EVP~\ref{evp_mu_closed} are further necessary
   to address the roles of $\mu$ and $d/R$.
These are made transparent when Equation~\eqref{eq_modal_EVP_govE} is transformed
   into a \Schrod\ form with standard techniques employed by, e.g.,
   \citetalias{1996ApJ...472..398B} and \citetalias{2015ApJ...810...87L},
   the result being
\begin{equation}
\label{eq_EVP_schro}
\begin{split}
&  \dispfrac{d^2 \Phi}{d r^2} 
 + Q(r) \Phi
 = 0, 	
\\ 
&  Q(r)
 = \dispfrac{\omega^2 - V(r)}{\va^2(r)}.
\end{split}
\end{equation}
Here $\Phi(r) = r^{1/2} \breve{v}(r)$ defines some ``wave function'', while 
\begin{equation}
   V(r) = \va^2(r)	\left(k^2 +\dispfrac{3}{4 r^2}\right)
   \label{eq_EVP_schro_Vpot}
\end{equation}   
   defines the potential. 
Three properties then ensue.
One, $\omgl$ is bound to exceed $k\vai$ regardless of
   $\mu$ or $d/R$ (see Appendix~\ref{sec_app_EVP_highfreq}). 
Two, high-frequency modes with $\omgl \gg k\vae$ are permitted regardless of
   $\mu$ or $d/R$, and they follow the relation $\omgl \approx l \pi\vae/d$ 
   when $d/R$ is sufficiently large
   (also see Appendix~\ref{sec_app_EVP_highfreq}).    
Three, the spatial behavior of mode functions $\breve{v}(r)$ is determined
   by the sign of $Q(r)$.
We therefore classify the modes into two categories, 
   labeling those with $Q(d) >0$ ($Q(d) <0$) as ``oscillatory'' (``evanescent'').
For the ease of description,   
   we see this classification scheme as applicable only to closed systems, 
   but adopt the viewpoint that such terms as ``trapped'' versus ``leaky''
   (e.g., \citetalias{1983SoPh...88..179E}; \citealt{1986SoPh..103..277C})
   or ``proper'' versus ``improper'' 
   (\citetalias{2015ApJ...806...56O}; also \citealt{2014ApJ...789...48O})
   apply to open systems. 
Evidently, an evanescent mode becomes a trapped mode when $d$
   goes infinite.
Its frequency $\omgl$ is therefore $d/R$-independent
   for sufficiently large $d/R$, and is bound to be lower than 
   the ``critical frequency'' $\omgcrit = k \vae$ because 
   $V(r) \to k^2 \vae^2$ when $r \to \infty$. 
Now specialize to our chosen set $[\rhoi/\rhoe, kR] = [2.25, \pi/15]$. 
Evanescent modes are possible only when $\mu <2$.
All modes are oscillatory for $\mu \ge 2$, the 
   frequency $\omega_l$ for any $l$ always higher than $\omgcrit$ and $d/R$-dependent.

\subsection{Numerical Results} 
\subsubsection{Frequency Distribution of Modal Contributions}
\begin{figure}
\centering
 \includegraphics[width=.7\columnwidth]{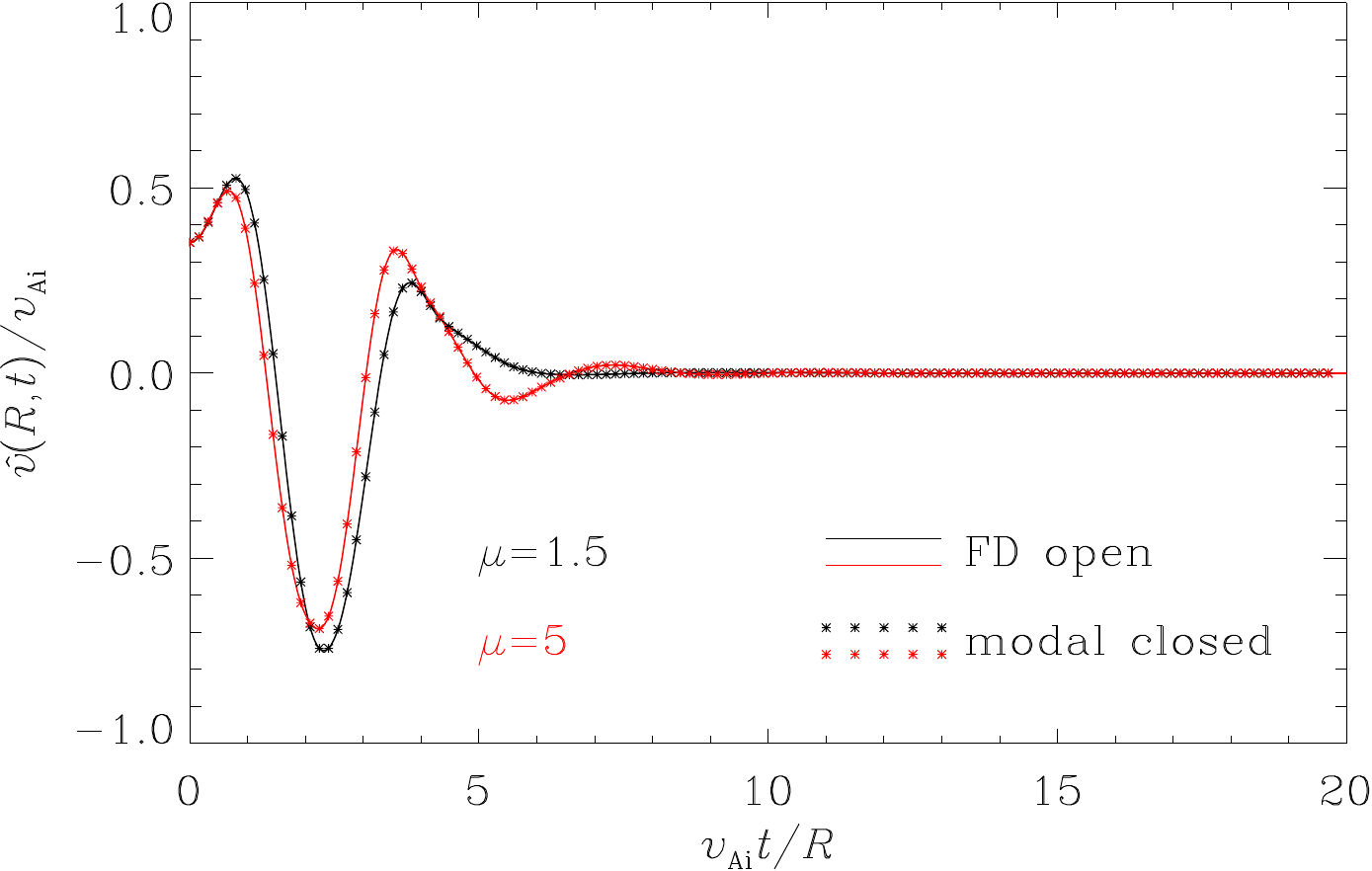}
 \caption{
 Temporal evolution of the radial speed at $r=R$ as found by solving
     IVP~\ref{ivp_mu_open} for two steepness parameters,
     one being $\mu=1.5$ (the black curves and symbols)
     and the other being $\mu=5$ (red).
 All computations pertain to the combination 
     $[\rhoi/\rhoe, kR, \Lambda/R] = [2.25, \pi/15, 4]$.
 Two independent methods are adopted to solve IVP~\ref{ivp_mu_open}.
 The finite-difference solutions are represented by the solid curves,
     labeled ``FD open" because this approach directly applies
     to a radially open system.
 The modal solutions are given by the asterisks,
     labeled ``modal closed'' because the solutions are based 
     on eigenmodes on a closed domain
     (see Equation~\eqref{eq_modal_formalSol}).
 A domain of size $d = 50~R$ is employed here.  
 }
 \label{fig_vtdep_FD_vs_modal}
\end{figure}

\begin{figure}
\centering
 \includegraphics[width=.8\columnwidth]{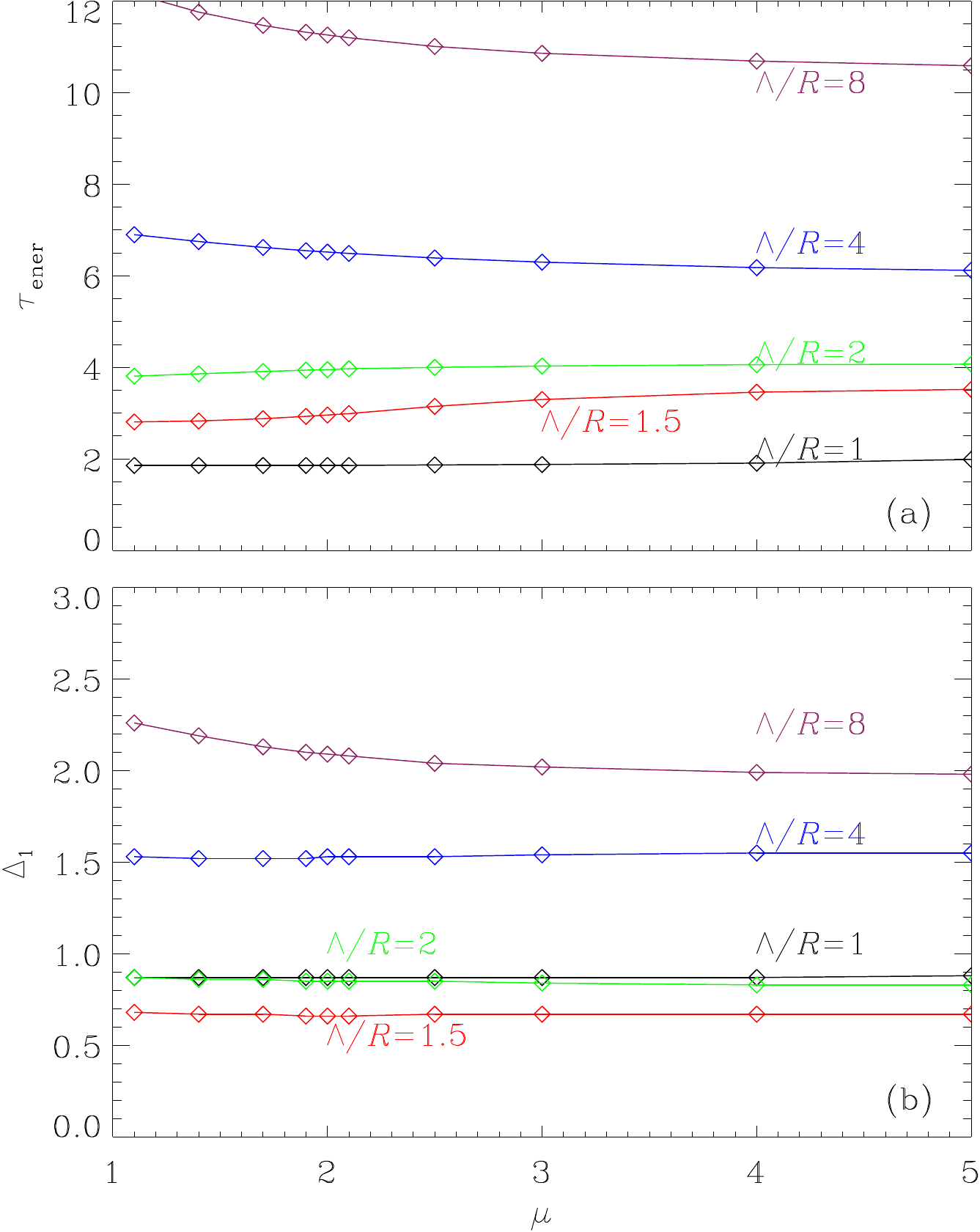}
 \caption{
 Dependencies of (a) $\tauener$ and (b) $\Delta_1$ on the 
     steepness parameter $\mu$ for a number of values
     of the spatial extent of the initial perturbation ($\Lambda$)
     as labeled. 
 Here $\tauener$ denotes the time at which the total wave energy
     in the cylindrical volume laterally bounded by $r=\Lambda$
     drops from its initial value by a factor of ${\rm e}^{4}\approx 55$.     
 Furthermore, $\Delta_1$ denotes the temporal spacing between
     the first two extrema in the $\hat{v}(R, t)$ profile. 
 All results are found by solving IVP~\ref{ivp_mu_open} with the finite-difference approach,
     and the combination 
     $[\rhoi/\rhoe, kR]$ is fixed at $[2.25, \pi/15]$.
 }
 \label{fig_survey_mu_Lambda}
\end{figure}

\begin{figure}
\centering
 \includegraphics[width=.8\columnwidth]{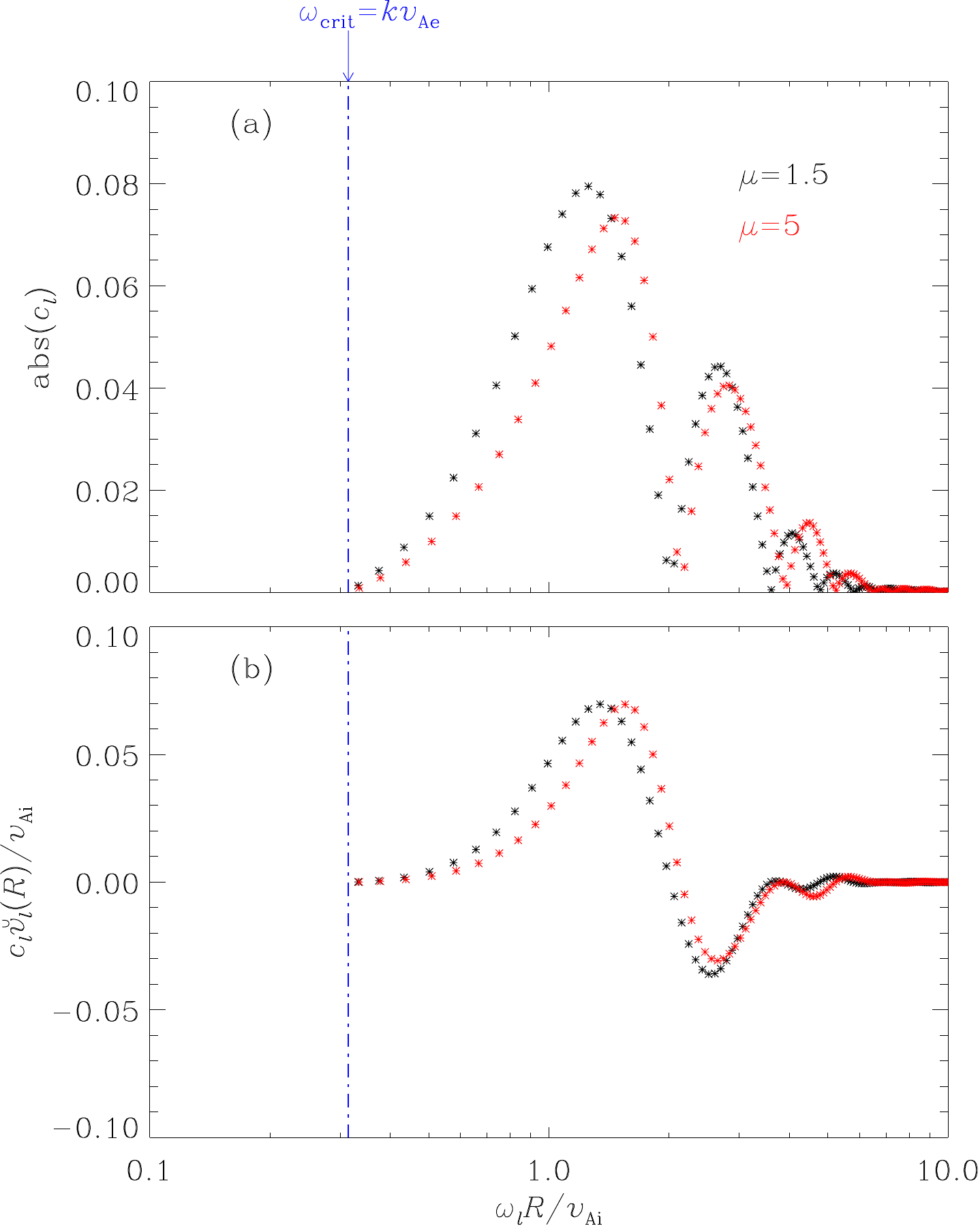}
 \caption{
Frequency-dependencies of 
    the contributions of individual modes 
    to the modal solutions 
    for a fixed combination $[\rhoi/\rhoe, kR, \Lambda/R, d/R] = [2.25, \pi/15, 4, 50]$.
 Two steepness parameters are examined, one being
    $\mu=1.5$ (the black asterisks) and the other being 
    $\mu=5$ (red).      
 Plotted are (a) the modulus of the position-independent coefficient $c_l$,
    and (b) the specific contribution $c_l \breve{v}_l (r)$ evaluated at $r=R$. 
 The critical frequency $\omgcrit = k \vae$ is marked 
    by the vertical dash-dotted lines for reference.
 }
 \label{fig_ME_modal_contrib}
\end{figure}

We fix the domain size to be $d=50~R$ in this subsection.
As shown by Figure~\ref{fig_vtdep_FD_vs_modal}, 
   the modal solutions thus constructed agree with their FD counterparts.
The dependencies on
   the mode frequency ($\omega_l$) of the contributions of individual modes
   are then shown in Figure~\ref{fig_ME_modal_contrib},
   where the modes are represented by the asterisks
   and different colors are adopted to discriminate
   between the two steepness parameters. 
The critical frequency $\omgcrit = k \vae$ is plotted 
   by the vertical dash-dotted lines for reference.
By examining $|c_l|$, Figure~\ref{fig_ME_modal_contrib}a 
   overviews the gross contributions
   from individual modes, where by ``gross'' we mean that $c_l$ is position-independent
   (Equation~\eqref{eq_modal_coef}).
For $\mu=1.5$ (the black symbols) and $\mu=5$ (red) alike, 
   the $\omega_l$-dependence of $|c_l|$ features a number of peaks whose
   magnitude weakens with $\omega_l$.
The modal contributions at the specific location $r=R$ are plotted 
   in Figure~\ref{fig_ME_modal_contrib}b, from which one sees that
   the spatial dependence of $\breve{v}_l (r)$ makes the contributions
   from modes with $\omega_{l} \gtrsim 3.5~\vai/R$ less pronounced
   than expected with Figure~\ref{fig_ME_modal_contrib}a.
Regardless, the point is that $\vhatd(R, t)$ is dominated by modes 
   with $\omega_{l}$ higher than but not far higher than $\vai/R$, 
   in agreement with the periodicities found in Figure~\ref{fig_vtdep_FD_vs_modal}.
More importantly, the lowest mode frequency $\omega_1$ exceeds $\omgcrit$, meaning that
   all modes are oscillatory. 
This is true not only for $\mu=5$ but also for $\mu=1.5$, despite that
   trapped modes are bound to appear in the latter case for a truly open system.

\subsubsection{Dependence of Modal Structure on Domain Size} 
This subsection examines how the modal structure depends 
    on the dimensionless domain size $d/R$,
    the reasons for doing which are twofold. 
Firstly, that trapped modes appear for an open system when $\mu <2$ 
    was found on solid mathematical grounds by \citetalias{2015ApJ...810...87L}.
One naturally argues that $d/R$ in Figure~\ref{fig_ME_modal_contrib}
    is not large enough for an evanescent mode to appear. 
However, the modal solution~\eqref{eq_modal_formalSol} does not depend on $d/R$
    within the timeframe explicitly given there.
Figure~\ref{fig_ME_modal_contrib} then indicates that 
    the contribution from evanescent modes is negligible    
    even if $d/R$ is larger still. 
We will quantify how negligible this contribution is.        
Secondly, to our knowledge, the only study involving
    EVP~\ref{evp_mu_closed} was conducted by \citetalias{1996ApJ...472..398B} for a step profile.
However, the $d/R$-dependence was not of interest and hence only briefly mentioned
    in Figure~2 therein.
We analytically examine this dependence in some detail for step and continuous
    profiles in Appendices \ref{sec_app_step} and \ref{sec_app_EVP_highfreq},
    respectively. 
The analytical results in turn help better quantify the modal behavior
    that we find numerically and present in the main text. 
    
We start by examining the $d/R$-dependence of the modal contributions
    to $\vhatd(r, t)$.
For this purpose we rewrite the modal solution~\eqref{eq_modal_formalSol} as
\begin{equation}
  \vhatd (r, t)
= \sum\limits_{\omega_{l} < k \vae} 
      c_l  \breve{v}_l(r) \cos(\omega_l t)
 +\sum\limits_{\omega_{l} > k \vae}  
      S_l \breve{v}_l(r) \cos(\omega_l t) \Delta\omega_{l},
  \label{eq_modal_Sol_SlInt}
\end{equation}
    where the second summation accommodates 
    all oscillatory modes whereas the first collects
    evanescent ones. 
The frequency spacing    
\begin{equation}
\Delta\omgl = \omega_{l+1}-\omega_{l}
	\label{eq_def_freqSpacing}
\end{equation}
    is relevant only for oscillatory modes in Equation~\eqref{eq_modal_Sol_SlInt}
    but actually defined for all $l$.
We further view the combination $S_l \breve{v}_l (r)$ as some 
    local ``spectral density'', with $S_l$ defined by       
\begin{eqnarray}
  S_l
= 
  \dispfrac{c_l}{\omega_{l+1}-\omega_{l}}.
    \label{eq_def_SpecDen}  	   
\end{eqnarray}    

\begin{figure}
\centering
 \includegraphics[width=.7\columnwidth]{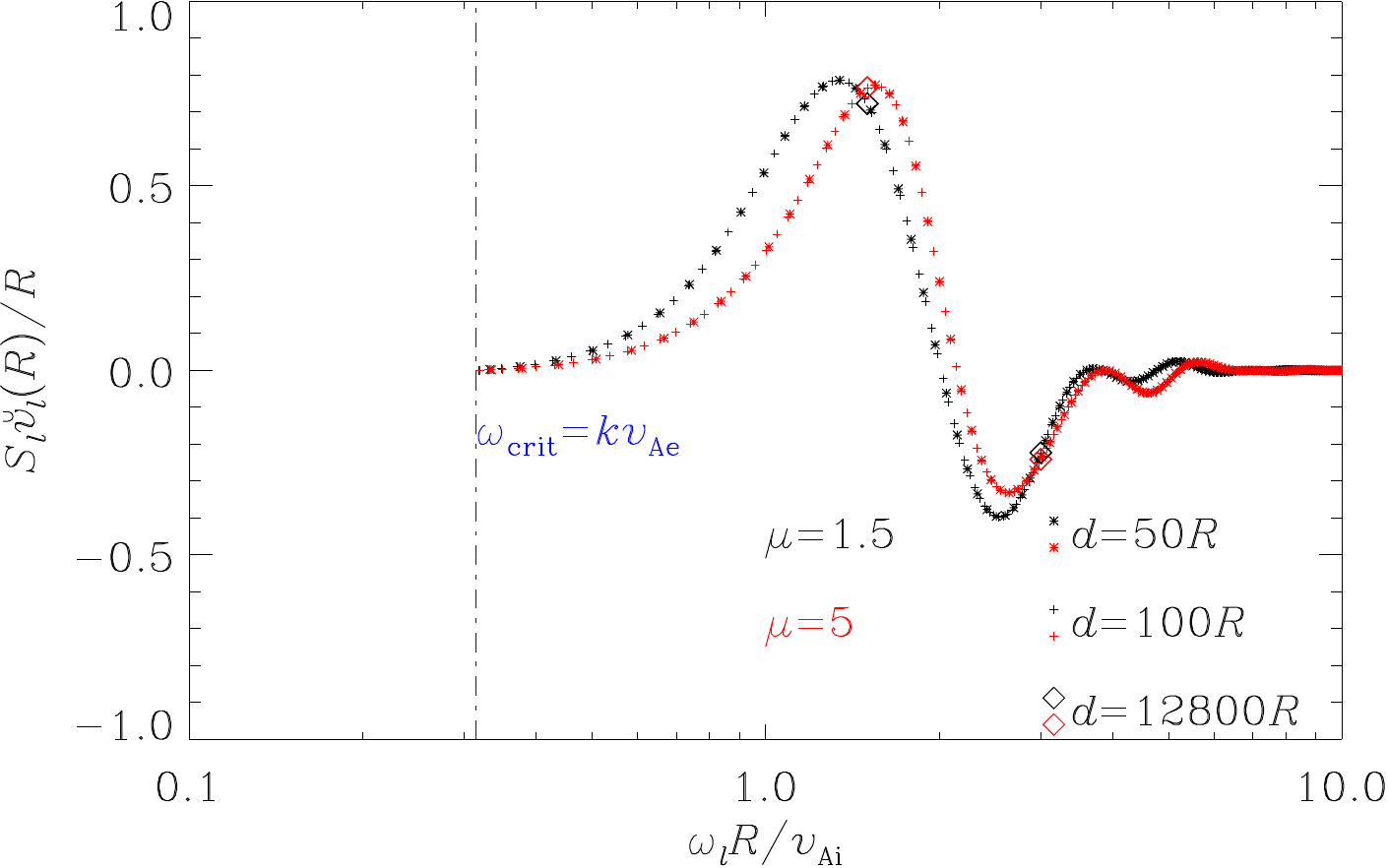}
 \caption{
Frequency-dependencies of the spectral density $S_l \breve{v}_l(r)$
    evaluated at $r=R$ involved in the modal solutions 
    for a fixed combination $[\rhoi/\rhoe, kR, \Lambda/R] = [2.25, \pi/15, 4]$.
Two steepness parameters are examined, one being
    $\mu=1.5$ (the black symbols) and the other being 
    $\mu=5$ (red).
The modal solutions are based on eigenmodes on a closed domain, for which 
    three different sizes ($d$) are experimented with.
Note that only two small subsets of modes are presented for the domain with
    $d/R = 12800$.
See text for details. 
 }
 \label{fig_ME_SpecDen_atR}
\end{figure}

Why is the modal solution decomposed in such
    a simple but cumbersome way?
We choose to leave a detailed answer until later,
    and for now stress only that it does not make sense to compare
    the frequency-dependencies of $c_l \breve{v}_l$ at a specific location
    between different values of $d/R$ because of the 
    $d/R$-dependence of $\omgl$.  
This is made more specific by Figure~\ref{fig_ME_SpecDen_atR},
    where we specialize to $r=R$ and display the $\omgl$-dependencies
    of the local spectral densities $S_l \breve{v}_l(R)$ 
    for both $\mu=1.5$ (the black symbols) and $\mu=5$ (red).       
Two domain sizes are examined, with the result for $d/R=50$ ($100$)
	represented by the asterisks (pluses).
For both values of $\mu$, 	
    one sees that the spectral densities for the two values of $d/R$
    outline exactly the same curve, even though
    the mode frequencies are 
    more closely spaced for $d/R = 100$ than for $d/R = 50$.
One further sees that all modes remain oscillatory for $\mu=1.5$ even when $d/R=100$.
In fact, an evanescent mode appears for this $\mu$ only when $d/R \gtrsim 4000$, 
    a remarkably large value that makes
    it numerically formidable to compute the necessary set of modes
    to yield a further $\omgl$-dependence
    of $S_l \breve{v}_l(R)$. 
We rather arbitrarily choose a $d/R=12800$ and 
    compute only two small subsets for each $\mu$, 
    the first (second) comprising those five modes
    with $\omgl$ just exceeding $1.5~\vai/R$ ($3~\vai/R$).
The corresponding spectral densities are plotted by
    the diamonds.
For each $\mu$, one then expects to see ten but actually can discern
    only two diamonds because the mode frequencies in each subset 
    are too close to tell apart. 
Regardless, the diamonds lie exactly on the curve outlined by the result
    for any smaller $d/R$, reinforcing the insignificance
    of evanescent modes despite that one such mode does exist.
To be precise, $c_1 \breve{v}_1 (R)$      
    for $\mu=1.5$ evaluates to $3.51\times 10^{-11}$
    in units of $\vai$ or rather in units of
    the magnitude of the initial perturbation.
    
\begin{figure}
\centering
 \includegraphics[width=.6\columnwidth]{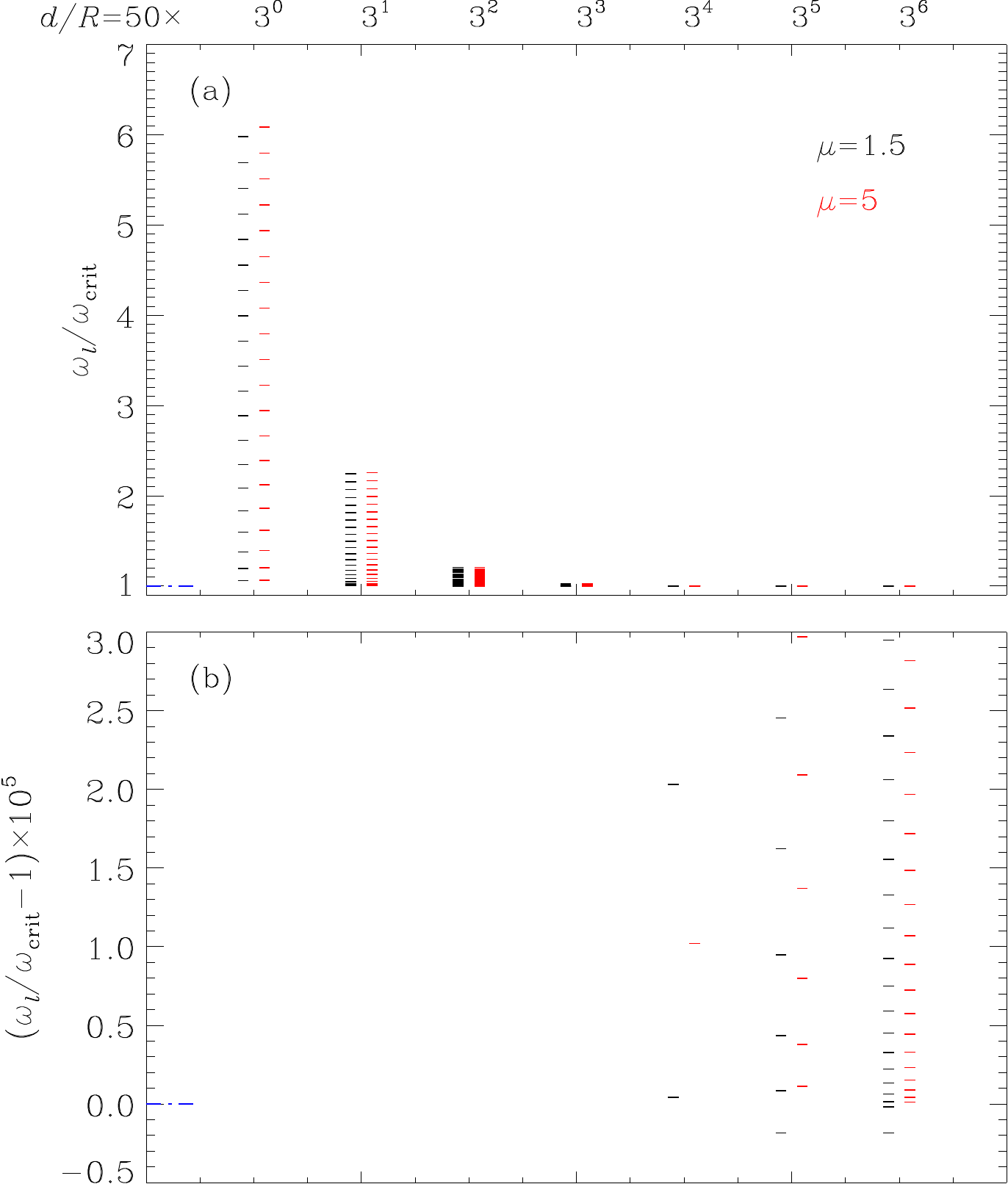}
 \caption{
 Eigenfrequency diagrams (``level schemes'') for the first $20$ modes
     found by solving
     EVP~\ref{evp_mu_closed} for a fixed combination $[\rhoi/\rhoe, kR] = [2.25, \pi/15]$.
 Two steepness parameters are examined, one being
	 $\mu=1.5$ (the black ticks) and the other being 
	 $\mu=5$ (red).
 A number of domain sizes ($d$) are examined as labeled,
     and the mode frequencies with a given $\mu$ are represented by
     the horizontal ticks stacked vertically for a given $d$.
 Plotted are (a) the mode frequencies in units of $\omgcrit = k\vae$,
     and (b) the fractional difference of mode frequencies from $\omgcrit$. 
 Note that this fractional difference is multiplied by 
     a factor of $10^5$.      
 }
 \label{fig_ME_level_scheme}
\end{figure}

We focus on how the modal structure varies when the domain size varies.
For a sequence of $d/R$ as labeled, Figure~\ref{fig_ME_level_scheme}a shows
    the frequencies ($\omgl$) of the first $20$ modes
    as horizontal ticks stacked vertically at a given $d/R$, 
    with the results for $\mu=1.5$ and $\mu=5$ differentiated by the black
    and red colors.
Note that in this ``level scheme'', $\omgl$ is measured in units
    of the critical frequency ($\omgcrit$),
    and the horizontal dash-dotted line marks $\omgl = \omgcrit$. 
Two features then follow.
By examining the case with $d/R = 50$ for either value of $\mu$,
    one sees that the frequency spacing
    $\Delta \omgl$ becomes increasingly uniform with increasing $l$.
As detailed in Appendices~\ref{sec_app_sub_step_modalclosed} and
    \ref{sec_app_EVP_highfreq}, this feature derives from
    the fact that for any $\mu$ at a sufficiently large $d/R$, 
    the mode frequency $\omgl$ for large enough $l$
    can be approximated by $l \pi \vae/d$ (Equation~\eqref{eq_app_high_endres}).
Slightly surprising is that this approximation is accurate to better than
    $5\%$ for both $\mu$ when $l$ merely exceeds $11$, despite that 
    $l$ is nominally required to be $\gg (d/R)/\pi \approx 15.9$
    for Equation~\eqref{eq_app_high_endres} to hold.
While not shown, we find that $\omgl$ may be approximated by $l \pi \vae/d$
    for $l$ beyond its nominal range of validity at other values 
    of $d/R$ as well.
Regardless, by ``feature~1'' we refer to the $\mu$-independent fact
    that $\omgl \propto l \vae/d$
    and hence $\Delta\omgl \propto \vae/d$ at large $l$ and large $d/R$.       
Feature~2, on the other hand, concerns the modes with $\omgl$
    that differs little from $\omgcrit$.
This turns out to be difficult to examine with Figure~\ref{fig_ME_level_scheme}a,
    because the mode frequencies become increasingly packed when $d/R$ increases.
In fact, $\Delta\omgl$ eventually becomes so small that
    we choose to exaggerate the fractional difference $\delta_l = \omgl/\omgcrit-1$
    by a factor of $10^5$ in Figure~\ref{fig_ME_level_scheme}b, 
    where the horizontal dash-dotted line again marks $\omgl = \omgcrit$.
Now one sees that $\omgl$ consistently exceeds $\omgcrit$ for $\mu=5$, 
    meaning that the modes are consistently oscillatory.
When $\mu=1.5$, however, the first mode shows up as
    an evanescent mode for $d/R = 50 \times 3^5 = 12150$, 
    and so does the second mode at the even larger $d/R$.

\begin{figure}
\centering
 \includegraphics[width=.7\columnwidth]{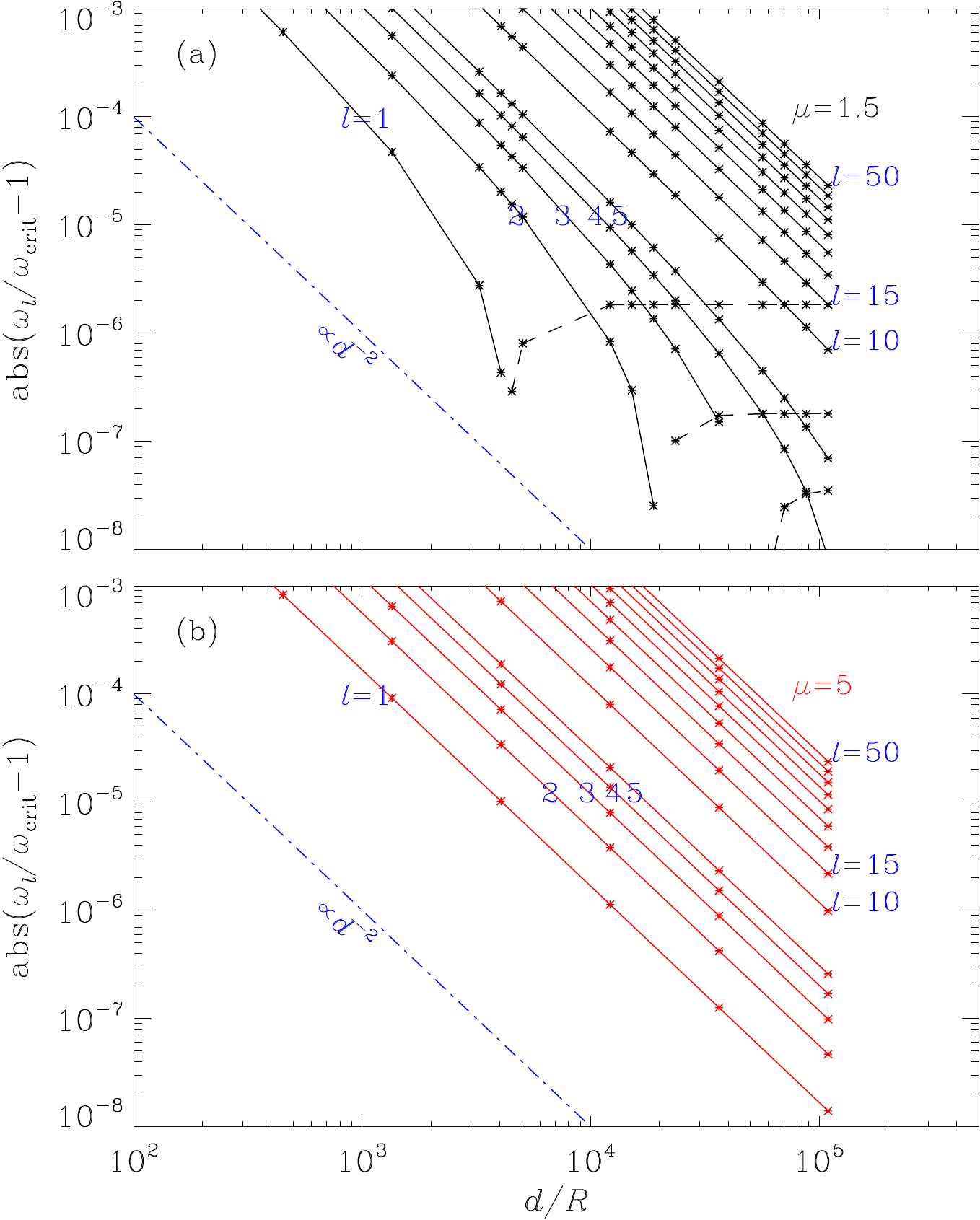}
 \caption{
Dependencies on the dimensionless domain size ($d/R$) of the   	
   modulus of the fractional difference  
   $\delta_l = \omega_l/\omgcrit-1$ 
   as found by solving EVP~\ref{evp_mu_closed}
   for a steepness parameter being 
   (a) $\mu = 1.5$ and (b) $\mu = 5$. 
The combination $[\rhoi/\rhoe, kR]$ is fixed at $[2.25, \pi/15]$.
For each pair $[\mu, d/R]$, the first five modes are always presented, while
   the rest are evenly sampled with a step of five in $l$ when $l$
   ranges from $10$ to $50$. 
In addition, $|\delta_l|$ for a given $l$ is connected 
   by a solid (dashed) curve when $\delta_l$ is positive (negative). 
The blue dash-dotted line represents a $1/d^2$-dependence for comparison.
See text for details.      
}
 \label{fig_modStr_fracDiff}
\end{figure}

\begin{figure}
\centering
 \includegraphics[width=.8\columnwidth]{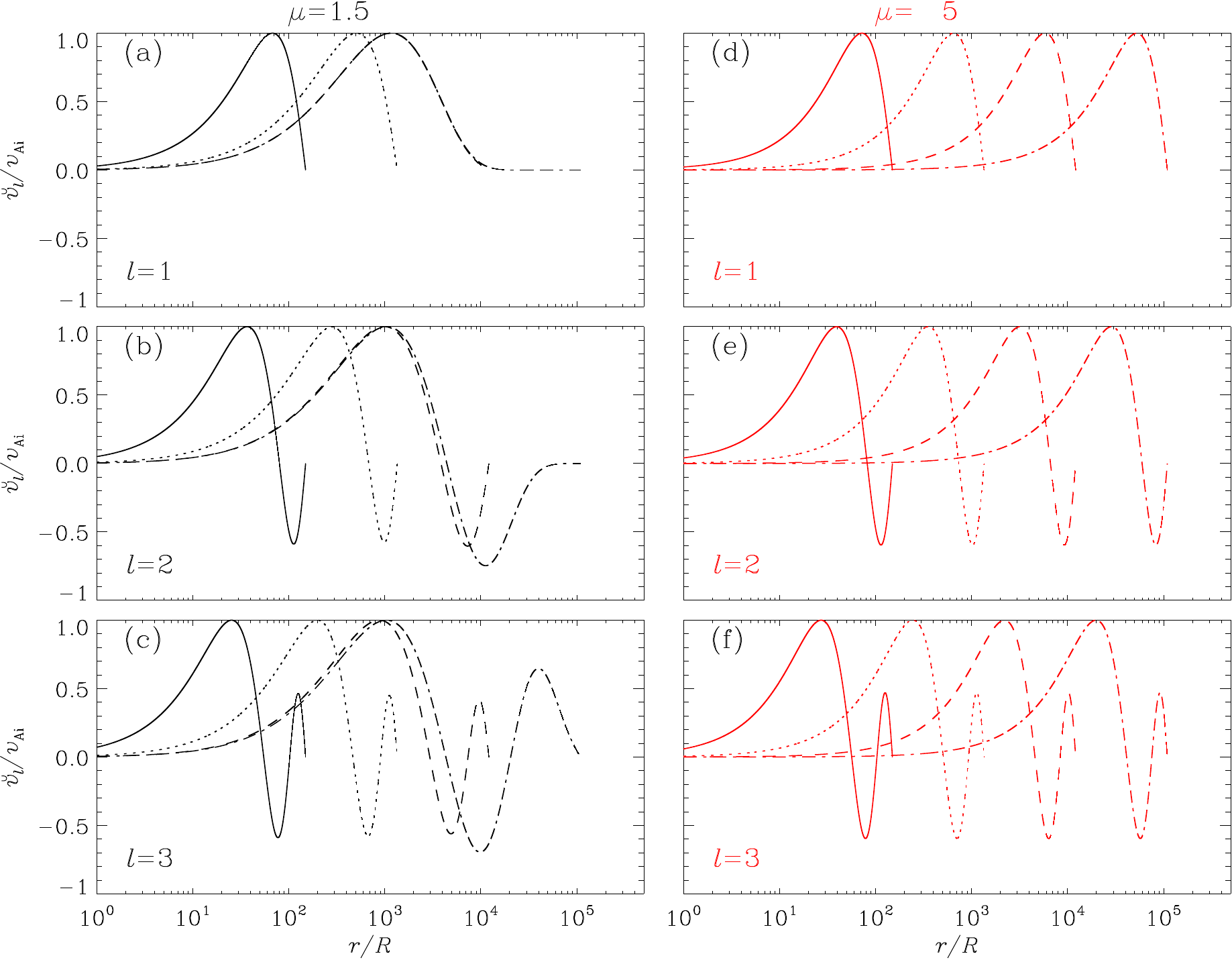}
 \caption{
Radial profiles of the first three eigenfunctions as found by solving
    EVP~\ref{evp_mu_closed} on a variety of domains differentiated 
    by the linestyles.
 The combination $[\rhoi/\rhoe, kR]$ is fixed at $[2.25, \pi/15]$.
 Two steepness parameters are examined, 
              one being $\mu = 1.5$ (the left column)
    and the other being $\mu = 5$ (right).
 The eigenfunctions are arbitrarily rescaled to better visualize
    the differences between different domain sizes.
 }
 \label{fig_modStr_eigFunc}
\end{figure}

Figure~\ref{fig_modStr_fracDiff} further examines the modes
    with $\omgl$ close to $\omgcrit$
    for (a) $\mu=1.5$ and (b) $\mu=5$.
Here the modulus of $\delta_l = \omgl/\omgcrit-1$ for a given $l$
    is displayed as a function of $d/R$ by the solid (dashed)
    curves when $\delta_l$ is positive (negative).
Among the $50$ modes examined, one mode out of five is plotted when $l$ ranges
    from $10$ to $50$, whereas all the first five are presented.
For reference, the eigenfunctions $\breve{v}_l$ of the first three modes
    are given by Figure~\ref{fig_modStr_eigFunc} for $\mu=1.5$ (the left column)
    and $\mu=5$ (right). 
A number of $d/R$ are examined and can be directly read from the figure. 
Examine the case with $\mu=5$ first.
Figure~\ref{fig_modStr_fracDiff}b indicates that $\delta_l$ is 
    positive for all modes, the oscillatory nature of which is made clearer
    by the spatial behavior of $\breve{v}_l$ in Figure~\ref{fig_modStr_eigFunc}.
Furthermore, $\delta_l$ for all $l$ follows a $1/d^{2}$-dependence
    shown by the blue dash-dotted curve in Figure~\ref{fig_modStr_fracDiff}b.
We note that this $1/d^{2}$-dependence is not empirically found but inspired
    by the analytical behavior of $\delta_l$ for step density profiles
    when $\delta_l \ll 1$ (see Equation~\eqref{eq_step_closed_dltvalid} for details).
That this dependence applies to the case $\mu=5$ reinforces the notion that
    the mode behavior is qualitatively similar when $\mu >2$.         
Now move on to the more interesting case where $\mu=1.5$.
For any of the first three modes, 
    Figure~\ref{fig_modStr_fracDiff}a indicates 
    a transition from an oscillatory to an evanescent mode as evidenced
    by the change of the sign of $\delta_l$ at some critical $\dcl$.
When $d/R$ becomes larger still, $\delta_l$ and hence $\omgl$ become independent
    of $d/R$.
In addition, $\dcl$ is seen to increase with $l$, a feature that can be readily
    understood with the left column of Figure~\ref{fig_modStr_eigFunc}.
Let $D_l$ denote the spatial extent of the eigenfunction of an evanescent mode,
    meaning mathematically that 
    {$Q(r)$ becomes negative when $r>D_l$}
     (see 
    Equations~\eqref{eq_EVP_schro} and \eqref{eq_EVP_schro_Vpot}).
When multiple evanescent modes exist on a sufficiently large domain,
    their frequencies are necessarily such that $\omega_1 < \omega_2 < \cdots$
    because the entire set $\{\omgl\}$ is a monotonically increasing sequence
    with respect to $l$.
On the other hand, it can readily shown that the potential $V(r)$ eventually
    approaches $k^2 \vae^2$ from below (above) when $\mu < 2$ ($\mu > 2$).
It then follows that $D_1 < D_2 < \cdots$.
Consequently, $(d/R)_{\rm cri, 1}$ is necessarily smaller than 
    $(d/R)_{\rm cri, 2}$ for the domain to accommodate
    the diminishing portion of the eigenfunction of the second mode.
In fact, the sequence ${\dcl}$ necessarily increases monotonically
    with respect to $l$. 
Figure~\ref{fig_modStr_fracDiff}a further indicates that 
	at sufficiently large $d/R$, the oscillatory modes for $\mu=1.5$ 
	remain characterized
    by $\delta_l \propto 1/d^2$ unless $\delta_l$ is extremely small.        

With the aid of Equation~\eqref{eq_modal_Sol_SlInt}, 
    we now offer some general remarks on the 
    dependencies on the steepness parameter $\mu$ of
    both the modal behavior on a closed domain with large $d/R$
    and the solution to IVP~\ref{ivp_mu_open}. 
However, we choose to focus on the chosen
    $[\rhoi/\rhoe, kR, \Lambda/R] = [2.25, \pi/15, 4]$ to avoid
    this manuscript becoming even longer. 
When $\mu <2$, more and more evanescent modes appear when 
    $d/R$ increases.
With the exception of the first several, the oscillatory modes are such that
    their frequency spacing $\Delta \omgl$ starts with a $1/d^2$-dependence
    before eventually settling to a $1/d$-dependence.
All modes are oscillatory when $\mu >2$, and the corresponding $\Delta \omgl$
    simply transitions from a $1/d^2$-dependence for small $l$
    to a $1/d$-dependence for large $l$.
Now focus on the two values of $\mu$ that we have adopted.
Recall that evanescent modes are irrelevant when $\mu = 5$,
    and make no contribution to the time-dependent solution when $\mu = 1.5$.
One therefore recognizes that only the second summation in
    Equation~\eqref{eq_modal_Sol_SlInt} matters.
What results from Equation~\eqref{eq_modal_Sol_SlInt} when $d/R$ increases
    is then an increasingly refined discretization of some Fourier integral 
    over a continuum of $\omega$ extending from $k\vae$ to infinity.
The relevant terms of this integral, 
    applicable to a truly open system ($d/R \to \infty$), was explicitly 
    worked out for $\mu=\infty$ by \citetalias{2015ApJ...806...56O}
    (see Appendix~\ref{sec_app_step_subOpen} for details).
Evidently, one byproduct of our modal approach on a closed domain is
    the numerical distribution
    in the $\omega-r$ space of the terms in the Fourier integral, 
    which cannot be expressed in closed-form for general $\mu$ to our knowledge.    
    
Supposing $\Lambda/R$ is adjustable in view of Figure~\ref{fig_survey_mu_Lambda},
    we move on to demonstrate a generic condition for evanescent modes 
    to be negligible.  
It suffices to adopt a truly open system, and
    consider the signal at a specific location such as $r=R$.
We start with the assumption that evanescent modes do not contribute,
    and deduce the condition that ensures this assumption.
Let $\tau$ denote the extent of the duration of interest, by which we mean that
    the signal becomes too weak to discern when $t>\tau$.     
Evidently the outermost edge travels to a distance of $D_\tau$.
Recall that the spatial extent of the eigenfunctions of the evanescent modes
    $D_l$ increases with $l$.         
One therefore deduces that evanescent modes are
    bound to be negligible when $D_\tau < D_1$.
On the other hand, $D_\tau$ is evidently lower than $\Lambda + \vae \tau$ because 
    the speed at which the outermost edge travels ($\va(r)$) 
    is consistently lower than $\vae$.
It then follows that evanescent modes can be neglected, provided
\begin{equation}
\label{eq_EvaNegligible}
    \Lambda + \vae \tau < D_1.
\end{equation}    
Equating $\tau$ to $\tau_{\rm ener}$ in Figure~\ref{fig_survey_mu_Lambda}, 
    one readily finds that the inequality holds for all the values of $\Lambda$
    examined therein, thereby explaining why the wave behavior 
    is solely characterized by dispersive propagation but shows no sign
    of wave trapping.
In fact, we can slightly generalize Equation~\eqref{eq_EvaNegligible}
    by supposing $\Lambda \gg R$ and adopting the worst-case scenario that
    $\tau$ is given by $t_{2+}$ in Equation~\eqref{eq_tmp1}.
We further neglect the deviation of $\va(r)$ from $\vae$, 
   meaning that $t_{1^{-}} (\Lambda/2 \to 0) \approx \Lambda/2\vae$
   and $t_{2^{+}} (0 \to \Lambda) \approx \Lambda/\vae$.
A rather safe estimate for $t_{2+}$ and hence $\tau$
    is then $3 \Lambda/2 \vae$.
The inequality \eqref{eq_EvaNegligible} therefore becomes 
\begin{equation}
\Lambda < \dispfrac{2 D_1}{5}.
\end{equation}
Note that the RHS evaluates to $\sim 4000~R$
    in view of Figure~\ref{fig_modStr_eigFunc},
    and further evaluates to $\sim 8 \times 10^6$~km
    if we quote an $R \sim 2000$~km from Figure~1 in
    \citet{2007ApJ...662L.119S} in keeping with
    the adopted $L/R=15$.
One therefore deduces that trapped modes are unlikely to be relevant
    in the temporal evolution of axial fundamental sausage modes in AR loops
    at least for the $[\rhoi/\rhoe, L/R]$ examined here.
We stress that trapped modes are allowed as eigensolutions on an open system 
    when $\mu<2$ as pointed out
    by \citetalias{2015ApJ...810...87L}.
Likewise, we stress that their distinct dispersive behavior
    relative to trapped modes in the canonical \citetalias{1983SoPh...88..179E}
    equilibrium is relevant when large values of $k$ are involved such as happens for impulsive
    sausage wavetrains in AR loops.
This latter point is clear if one contrasts Figure~8 with Figure~3 
    in \citet{2017ApJ...836....1Y}.
It is just that the existence of trapped modes for $\mu<2$ on an open system
    does not guarantee that they contribute to the temporal evolution of
    axial fundamentals.

\section{Summary}
\label{sec_conc}

Focusing on fast sausage modes (FSMs) in solar coronal loops, 
    this study aimed at examining   
    the consequences of some peculiar dispersive properties
    that may arise in an equilibrium 
    differing from \citet[][\citetalias{1983SoPh...88..179E}]{1983SoPh...88..179E} 
    only by replacing the step with a continuous density profile ($\rho_0(r)$). 
By ``peculiar'' we mean that FSMs 
    are not subject to cutoff axial wavenumbers when $\rho_0(r)$ outside
    the cylinder possesses a sufficiently shallow $r$-dependence,
    which was first
    recognized on firm mathematical grounds
    by \citet[][\citetalias{2015ApJ...810...87L}]{2015ApJ...810...87L}
    when analyzing the relevent eigenvalue problem (EVP) on a radially open system.
Two effects follow.
Firstly, FSMs may be trapped regardless of the axial wavenumber $k$
    and the density contrast $\rhoi/\rhoe$.
Secondly, long-wavelength trapped FSMs 
    are nearly dispersionless with their axial phase speeds  
    differing little from the external \Alf\ speed.
These two effects then led
	\citetalias{2015ApJ...810...87L} and \citet{2019MNRAS.488..660L}
	to deduce that fast sausage perturbations of observable quality
	may exist in active region (AR) loops and flare loops alike,
	with their periodicities characterized by the longitudinal rather 
	than the canonical transverse \Alf\ time.
If true, this deduction may substantially broaden the range of observed periodicities 
    that FSMs can account for, and therefore offer more seismological
    possibilities. 
    
We took efforts to make our scope as narrow as possible by addressing
    the question ``does the existence of trapped modes in EVPs on an open system
    guarantee that they play a role in determining the temporal behavior
    of sausage perturbations''?
To be specific, we chose to work in the framework of
    linearized, pressureless, ideal MHD, 
    and specialize to an ``outer-$\mu$'' density profile (Equation~\eqref{eq_rho_prof_outermu}).    
The solution to the relevant initial value problem
	on an open system
    (IVP~\ref{ivp_mu_open})
    is then determined by the dimensionless parameters
    $[\rhoi/\rhoe, \mu; kR; \Lambda/R]$, where $\mu$ characterizes
    the steepness of the density profile outside the nominal radius $R$,
    and $\Lambda$ represents the spatial extent of the initial perturbation. 
We focus on axial fundamentals in AR loops by
    taking $[\rhoi/\rhoe, kR] = [2.25, \pi/15]$ in view of the observational constraints
    from EUV measurements \citep{2004ApJ...600..458A,2007ApJ...662L.119S}.
We distinguish between the cases with $\mu < 2$ and $\mu>2$ because trapped modes 
    are present (absent) in the former (latter) for the chosen $[\rhoi/\rhoe, kR]$.
IVP~\ref{ivp_mu_open} is solved with both a direct finite-difference (FD) approach
    and a modal approach whereby the solution is expressed 
    as the superposition of eigenmodes for the 
    pertinent EVP on a closed domain (EVP~\ref{evp_mu_closed}).
The dimensionless domain size $d/R$ is involved in the latter approach,
    the evanescent modes in which are the counterparts of trapped modes on a truely open system.
Our findings can be summarized as follows. 

The answer to the question we laid out is ``No". 
We came to this conclusion primarily 
    because the FD solutions
    for a substantial range of $\mu$ and $\Lambda/R$
    are consistently characterized by some dispersive propagation
    but show no sign of wave trapping (Figure~\ref{fig_survey_mu_Lambda}).
In particular, the solutions show a smooth transition when $\mu$ crosses 
    the nominally critical value of two. 
With the modal approach, we showed that more and more evanescent modes appear
    when the domain size increases, thereby lending further support to 
    the recognition of \citetalias{2015ApJ...810...87L}.
However, even the shortest spatial extent of the evanescent eigenfunctions
    is well beyond the observationally reasonable range of
    the spatial extent of initial perturbations
    \footnote{
    Some subtleties arise as a result of our outer-$\mu$ formulation.
    Further computations are therefore conducted, the descriptions of which are
       nonetheless collected in Appendix~\ref{sec_app_FD_scale} to streamline
       the main text. 
    Our conclusion remains valid, namely that the existence of trapped modes in EVPs
       on an open system does not necessarily mean that they can show up in the evolution of the system in response to sausage-type perturbations. 
    }.
Consequently, the initial perturbations cannot impart a discernible
    fraction of energy to evanescent modes, which in turn means that
    these modes do not contribute to the temporal evolution of the system.

Before closing, we offer some further remarks on the influence on coronal FSMs
    due to the deviation of the equilibrium from \citetalias{1983SoPh...88..179E}.
We start by noting that the formulation of the transverse structuring 
    actually offers a mixed message for FSMs in terms of 
    their observational applications.
On the one hand, sausage-like perturbations are robust in the sense that
    they are permitted even when
    waveguides are not strictly axisymmetric as happens 
    for waveguides with, say, elliptic 
    \citep[e.g.,][]{2009A&A...494..295E,2021ApJ...912...50A} 
    or even irregular cross-sections \citep{Aldhafeeri21,2021ApJ...921L..17G}.
On the other hand, that the dispersive behavior of FSMs in
    non-\citetalias{1983SoPh...88..179E} equilibria
    may be qualitatively different does not necessarily mean that
    FSMs can be invoked to interpret a broader range of periodicities.
That said, one cannot rule out coronal FSMs as an interpretation
    for oscillations with periodicities on the order of
    the longitudinal \Alf\ time either.
Let us name only one possible equilibrium configuration where 
    essentially the only difference from \citetalias{1983SoPh...88..179E}
    is the introduction of a magnetically twisted boundary layer.
Radial fundamental FSMs may be trapped regardless of the axial
    wavenumber \citep[e.g.,][]{2012SoPh..280..153K,2018JASTP.175...49L},  
    the corresponding eigenfunctions possessing spatial scales
    that do not far exceed the cylinder radius
    \citep[e.g.,][]{2014SoPh..289.4069M,2021MNRAS.505.1878L}.
One readily deduces that these radial fundamental
    FSMs may indeed show up in reality, even though a definitive answer relies
    on a detailed study from the IVP perspective.

\acknowledgments
{We thank the referee for constructive comments.}
This research was supported by the 
    National Natural Science Foundation of China
    (41974200 and 11761141002).
We gratefully acknowledge ISSI-BJ for supporting the workshop
    on “Oscillatory Processes in Solar and Stellar Coronae”, 
    during which this study was partly initiated. 

\bibliographystyle{aasjournal}
\bibliography{seis_generic}

\appendix
\section{Standing Sausage Modes in Coronal Cylinders with Step Profiles}
\label{sec_app_step}    

\subsection{Solution to IVP~\ref{ivp_mu_open} in Terms of Eigenmods for an Open System}
\label{sec_app_step_subOpen} 
This subsection presents the formal solution to
    IVP~\ref{ivp_mu_open} expressed as the superposition
    of eigenmodes for a laterally unbounded system.
We closely follow the Fourier-integral-based approach that
    \citetalias{2015ApJ...806...56O} adopted to examine
	the two-dimensional (2D) propagation
    of sausage wavetrains in a system that is unbounded 
    in the axial direction as well.
The modal solution to our IVP~\ref{ivp_mu_open} is actually part 
    of their 2D solution.
Consequently, only slight
    revisions to Equation~(25) in \citetalias{2015ApJ...806...56O}
    are needed to ensure dimensional and notational 
    consistency.  
We choose to provide a minimal set
    of equations leading to the modal solution for two reasons.
One, some equations will find immediate applications
    to the pertinent EVP on a closed domain.
Two, the conceptual understanding embedded in 
    the formal solution to IVP~\ref{ivp_mu_open}
    should be informative for future studies on standing sausage perturbations
    in generic coronal structures.
To this end, we see the axial wavenumber ($k$)
    as given and arbitrary. 
        
The following notations are necessary.
We use $J_n$ and $Y_n$ to represent
   Bessel functions of the first and second kind, respectively.
Likewise, $I_n$ and $K_n$ denote modified Bessel functions of
   the first and second kind, respectively.
Only orders $n=0$ and $1$ are relevant.
In particular,    
   $j_{n, m}$ denotes the $m$-th zero of $J_n$ with $m=1, 2, \cdots$. 
The cutoff wavenumbers are then given by
\begin{eqnarray}
  k_{\rm cutoff, m} R 
= \dispfrac{j_{0, m}}{\sqrt{\rhoi/\rhoe-1}}. 
   		\label{eq_step_kcut} 
\end{eqnarray}            
Defining
\begin{equation}
\label{eq_step_defkikekappae}
\begin{split}
& \ki^2  
   =  \dispfrac{\omega^2 - k^2 \vai^2}{\vai^2}, \\
& \ke^2  
   =  \dispfrac{\omega^2 - k^2 \vae^2}{\vae^2}, \\
& \kappa_{\rm e}^2
   = -\dispfrac{\omega^2 - k^2 \vae^2}{\vae^2}
   = -\ke^2, 
\end{split}
\end{equation}    
    we note that $\omega$ always exceeds $k\vai$ 
    and hence $\ki^2$ are always positive.
       
The modal solution to IVP~\ref{ivp_mu_open}
    involves both proper and improper modes, which
    are discriminated by the sign of $\ke^2$. 
Proper modes ($\ke^2 <0$) are relevant when $k$ exceeds $k_{\rm cutoff, 1}$, 
    and correspond to a discrete set of
    frequencies. 
Let $j$ label a proper mode.
Its eigenfunction reads
\begin{eqnarray}
\breve{v}_j (r)=  
  \left\{
    \begin{array}{ll}
      -\dispfrac{\vai}{\ki R}      K_0(\kappae R) J_1(\ki r),   
		        & \quad 0 \le r \le R, 			\\[0.1cm]
       \dispfrac{\vai}{\kappae R}  J_0(\ki R)     K_1(\kappae r),   
				& \quad r > R.
    \end{array} 
 \right.
 \label{eq_step_open_vproper}
\end{eqnarray}        
Written in this form, Equation~\eqref{eq_step_open_vproper} ensures that
    $d\breve{v}_j(r)/dr$ is continuous. 
The mode frequency is then dictated by the continuity of $\breve{v}_j$ itself
    across $r=R$, which leads to the well-known dispersion relation
    (DR, e.g., \citetalias{1983SoPh...88..179E};
    also
     \citealt{1978SoPh...58..165M,1982SoPh...75....3S,1986SoPh..103..277C})
\begin{eqnarray}
   \ki     \dispfrac{J_0 (\ki     R)}{J_1 (\ki R)}
+  \kappae \dispfrac{K_0 (\kappae R)}{K_1 (\kappae R)}
   = 0~.
   \label{eq_step_open_properDR}
\end{eqnarray}    
Improper modes ($\ke^2 >0$) are relevant regardless of $k$, their frequencies
   continuously spanning the range $(k\vae, \infty)$.
The eigenfunction reads
\begin{eqnarray}
\breve{v}_\omega (r)=  
  \left\{
    \begin{array}{ll}
      -\vai \dispfrac{\ke^2 R}{\ki}  J_1(\ki r),   
		        & \quad 0 \le r \le R, 			\\[0.1cm]
      -\vai (\ke R) 
         \left[C_J J_1(\ke r) + C_Y Y_1 (\ke r) \right],   
				& \quad r > R,
    \end{array} 
 \right.
 \label{eq_step_open_vImproper}
\end{eqnarray}        
   where 
\begin{equation}
\label{eq_step_open_CJCY}           
\begin{split} 
& C_J = \dispfrac{\pi \ke R}{ 2\ki}  
      \left[
         -\ki J_0(\ki R) Y_1(\ke R) + \ke J_1(\ki R) Y_0(\ke R)
      \right], \\
& C_Y = \dispfrac{\pi \ke R}{ 2\ki}  
      \left[
         -\ke J_1(\ki R) J_0(\ke R) + \ki J_0(\ki R) J_1(\ke R)
      \right].
\end{split}
\end{equation}  
Both $\breve{v}_\omega$ and $d\breve{v}_\omega(r)/dr$ are ensured to be continuous.
The modal solution eventually reads
\begin{equation}
	\label{eq_step_open_SolAna}      
\begin{split}
&    \hat{v}(r, t)
   = \sum\limits_{j = 1}^{J} c_j     
          \breve{v}_j(r)      \cos(\omega_j t) 
   + \int_{k \vae}^{\infty}  S_\omega 
	      \breve{v}_\omega(r) \cos(\omega   t) d\omega,
\\
&   0< r < \infty, \quad 0< t < \infty,	      
\end{split}	
\end{equation}       
   where 
\begin{equation}
\label{eq_step_open_Sol_DefcjSomg}
\begin{split}
& c_j = 
      \dispfrac{\displaystyle \int_{0}^{\infty} u(r) \breve{v}_j(r)  
					 \rho_0(r) r dr}
               {\displaystyle\int_{0}^{\infty}        \breve{v}^2_j(r)
               	     \rho_0(r) r dr}, 	\\
& S_\omega =                	     
	  \dispfrac{\displaystyle \omega \int_{0}^{\infty} u(r)
	    			 \breve{v}_\omega(r) \rho_0(r) r dr}
	           {(\rhoe \vai^2 R^2) (\ke \vae)^2 (C_J^2+C_Y^2)}.
\end{split}
\end{equation} 

Let us summarize the steps to solve IVP~\ref{ivp_mu_open}
   given $\rhoi/\rhoe$, $k$, and $u(r)$.          
First of all, with Equation~\eqref{eq_step_kcut} one counts
    $J$, the number of cutoff wavenumbers that are smaller than $k$.
Evidently, $J=0$ if $k < k_{\rm cutoff, 1}$, making proper modes
    irrelevant.
Secondly, if $J \ge 1$, then for each allowed $j$ one evaluates its
    eigenfrequency $\omega_j$ and then its eigenfunction $\breve{v}_j$
    with Equations~\eqref{eq_step_open_properDR}
    and \eqref{eq_step_open_vproper}, respectively.
The contribution of the proper mode, $c_j$, can then be found with
    Equation~\eqref{eq_step_open_Sol_DefcjSomg}.          
Thirdly, with Equations~\eqref{eq_step_open_vImproper} 
    and \eqref{eq_step_open_CJCY} one evaluates the improper
    eigenfunction for any $\omega > k \vae$.         
The contribution from the improper mode, $S_\omega$,
    is then readily found with Equation~\eqref{eq_step_open_Sol_DefcjSomg}.

\subsection{Eigenmodes for a Closed System}
\label{sec_app_sub_step_modalclosed}
\begin{figure}
\centering
 \includegraphics[width=.8\columnwidth]{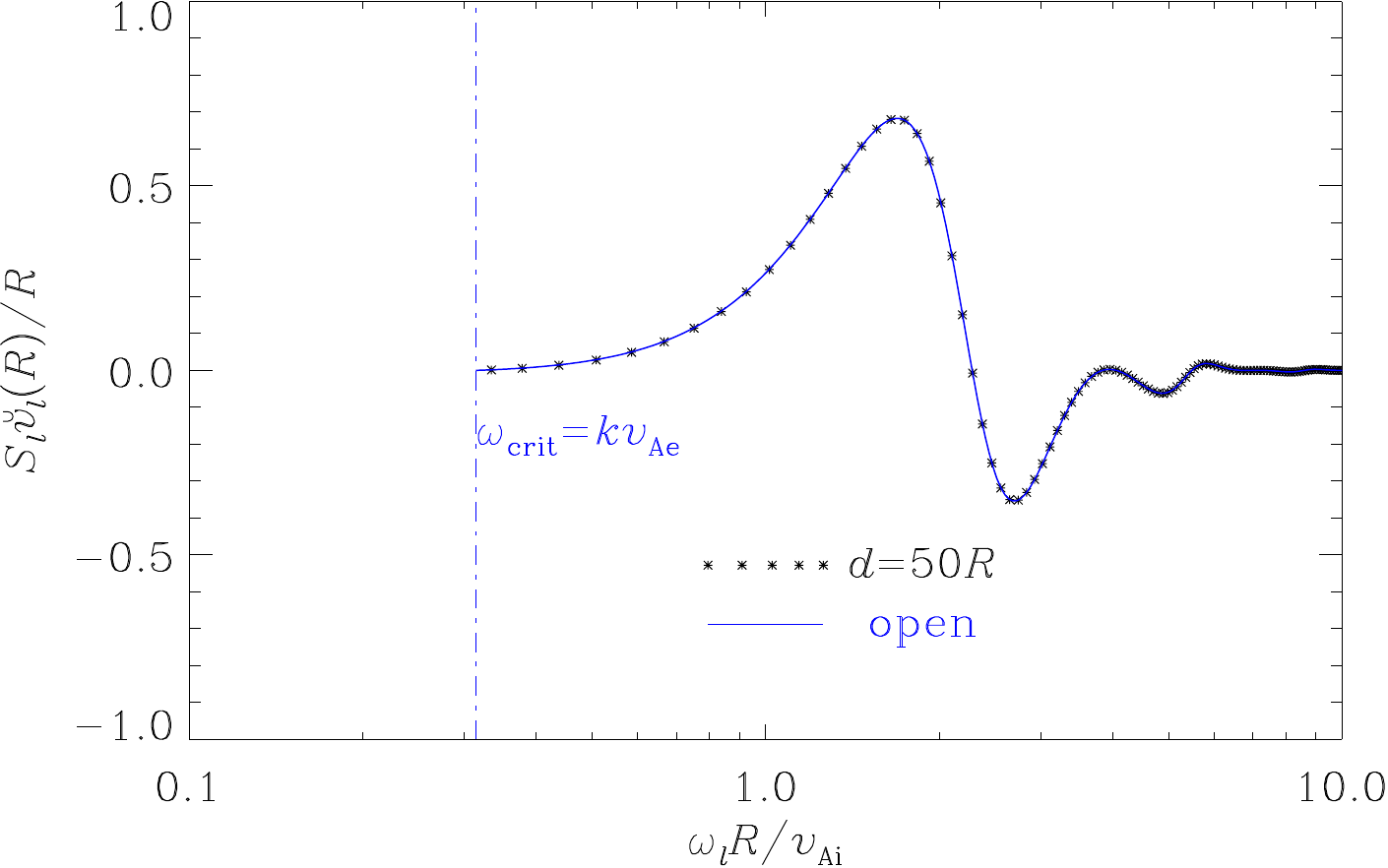}
 \caption{
Frequency-dependencies of spectral densities $S\breve{v}(r)$ 
   evaluated at $r=R$ as found by 
   solving IVP~\ref{ivp_mu_open}
   for a coronal cylinder with a step density profile.
The combination $[\rhoi/\rhoe, kR, \Lambda/R]$ is fixed at $[2.25, \pi/15, 4]$.   
The asterisks represent the discrete modes pertinent to EVP~\ref{evp_mu_closed}
   on a domain with $d/R = 50$,
   whereas the blue solid curve represents the continuum of improper eigenmodes 
   on a radially open system.
The vertical dash-dotted line marks the critical frequency 
   $\omgcrit = k \vae$.    
See text for details.    
 }
 \label{fig_step_modContri}
\end{figure}

This subsection examines some properties of the eigenmodes for a closed system,
    namely the solutions to EVP~\ref{evp_mu_closed} specialized
    to a step density profile. 
We start with a concrete example found for 
    $[\rhoi/\rhoe, kR] = [2.25, \pi/15]$ and an initial perturbation
    given by Equation~\eqref{eq_u} with $\Lambda/R = 4$.
IVP~\ref{ivp_mu_open} is solved with three independent methods.
The first, to be called ``modal open'', is based on eigenmodes on an open system,
    whereas the second (``modal closed'') is based on Equation~\eqref{eq_modal_formalSol}
    for a domain size $d = 50~R$.
The two sets of solutions agree exactly with each other, 
    and further agree
    with the solution found with the simpler finite-difference approach.
We choose not to present the comparison among the time-dependent solutions, 
    because a comparison between the frequency-dependencies of 
    the modal contributions seems more informative
    but is unavailable as far as we know.
Note that these contributions for the chosen $[\rhoi/\rhoe, kR]$
    are solely due to improper (oscillatory) modes
    in the ``modal open'' (``modal closed'') approach.  
With Equations~\eqref{eq_modal_Sol_SlInt} and \eqref{eq_step_open_Sol_DefcjSomg} 
    in mind, 
    Figure~\ref{fig_step_modContri} then specializes to $r=R$
    and compares the local spectral density $S_l \breve{v}_l(R)$
    from the ``modal closed'' approach (the asterisks)
    with $S_\omega \breve{v}_\omega(R)$ from the ``modal open'' approach (the blue solid curve).
It is reassuring to see that the solid curve threads exactly the symbols, meaning that 
    the continuum of improper modes is adequately resolved
    by the discrete oscillatory modes despite the rather modest
    domain size.

We now focus on the discrete modes themselves by recalling the discussions
    immediately following Equation~\eqref{eq_EVP_schro}.
Firstly, $\ki^2 >0$ is guaranteed because     
    $\omgl$ necessarily exceeds $k\vai$.
Secondly, the mode classification is eventually determined by how 
    $\omgl$ compares with $k\vae$, meaning that
    $\ke^2$ is positive (negative)
       	when a mode is oscillatory (evanescent).
Regardless, the eigenfunction $\breve{v}_l(r)$
	can be consistently	described by
    Equation~\eqref{eq_step_open_vImproper}, the reason being that
    $J_1 (\ke r)$ and $Y_1 (\ke r)$
    always form a numerically satisfactory pair 
    in the outer region ($R < r < d$)
    \footnote{The outer solution can be equivalently expressed 
       by the numerically satisfactory pair $I_1 (\kappae r)$
       and $K_1 (\kappae r)$ \citepalias[see][for details]{1996ApJ...472..398B}.
       It is just that $\kappae^2$ is positive (negative)
          for evanescent (oscillatory) modes.
     }.
   The requirement $\breve{v}_l (r=d) =0$ then gives a DR
       that governs the mode frequency $\omega_l$,
   \begin{eqnarray}
   C_J J_1 (\ke d) + C_Y Y_1 (\ke d) = 0.
       \label{eq_step_closed_DR}
   \end{eqnarray} 
We now specialize to the situation where
   $k < k_{\rm cutoff, 1}$ to better connect
   with the main text, the associated analytical progress
   being new to our knowledge.
    
Let us examine the analytical behavior of the modes with frequencies
	$\omega_l$ just above $\omgcrit = k\vae$.
Expressing $\omega_l$ as $k \vae (1+\delta)$ with $0<\delta \ll 1$, 
    one finds that $\ki^2 \approx k^2 (\rhoi/\rhoe -1)$
    and $\ke^2 \approx k^2 (2 \delta)$
    (see Equation~\eqref{eq_step_defkikekappae}).
Now suppose that $\ke R \ll 1$.
The approximate expressions of $J_n$ and $Y_n$ for small arguments
    then indicate that $Y_1 (\ke R) \sim 1/(\ke R)$
    is the most singular term in the coefficients $C_J$ and $C_Y$ (see Equation~\eqref{eq_step_open_CJCY}). 
The left-hand side (LHS) of the DR~\eqref{eq_step_closed_DR}
    is therefore dominated by the first term, meaning that 
    $\ke d \approx j_{1, l}$. 
In other words, 
\begin{eqnarray}
\delta =    \dispfrac{\omega_l}{k\vae}-1
    \approx \dispfrac{j_{1, l}^2}{2 (k d)^2}.
    \label{eq_step_closed_delta}
\end{eqnarray}
Given the assumptions $\delta \ll 1$ and $\ke R \ll 1$, 
    the modes in question are characterized by
\begin{equation}
\label{eq_step_closed_dltvalid}
\begin{split}
&         \dispfrac{\omgl}{k\vae} 
  \approx 1+\dispfrac{j_{1, l}^2}{2 (k d)^2}, \\
&  \mbox{provided}~
   d/R \gg j_{1, l}, \quad 
   (kd)^2 \gg j_{1, l}^2.  
\end{split}
\end{equation}    

Now consider high-frequency modes in a system
    with $d/R \gg 1$.
By ``high'' we assume that
\begin{eqnarray}
  \omega_l \gg \vae/R, \quad 
  \omega_l \gg k\vae.
  \label{eq_step_closed_omglargeAssump} 
\end{eqnarray}
It then follows from Equation~\eqref{eq_step_defkikekappae} 
    that $\ki R, \ke R \gg 1$
    and hence $\ki d, \ke d \gg 1$.
With the expressions
   for $J_n$ and $Y_n$ at large arguments,
   some algebra indicates that the DR~\eqref{eq_step_closed_DR}
   approximates to
\begin{eqnarray}
   \sin[\ke (d - R) + \phi_{\rm e}] = 0, 
   \label{eq_step_tmp1}
\end{eqnarray}   
   where $\phi_{\rm e}$ satisfies the relation 
\begin{eqnarray}
  \tan\phi_{\rm e} 
= \sqrt{\rhoe/\rhoi}
  \tan\left[
  		\sqrt{\rhoi/\rhoe} (\ke R) - \dispfrac{\pi}{4}
      \right].
      \label{eq_step_tmp2}
\end{eqnarray}
Note that Equation~\eqref{eq_step_tmp1} is implicit in $\omega_l$
   because $\omega_l$ is involved in Equation~\eqref{eq_step_tmp2}.
Note further that the range of $\phi_{\rm e}$ is not restricted 
   by Equation~\eqref{eq_step_tmp2} per se.
However, as can be verified a posteriori, $\phi_{\rm e}$
   is negligible to leading order, meaning that
   $\ke \approx l\pi/(d-R) \approx l\pi/d$. 
If desired, this solution can be plugged into 
   Equation~\eqref{eq_step_tmp2} to yield
   the first-order correction due to $\phi_{\rm e}$, resulting in
\begin{eqnarray}
\omega_{l} \approx 
   \left(l \pi - \phi_l\right)\dispfrac{\vae}{d-R}.
   \label{eq_step_closed_omgLarge}
\end{eqnarray}   
Here $\phi_l$ given by
\begin{eqnarray}
  \phi_l 
= \arctan\left[
	      \sqrt{\rhoe/\rhoi}
	      \tan\left(\sqrt{\rhoi/\rhoe} \dispfrac{l\pi}{d/R-1}
	      	       -\dispfrac{\pi}{4}
	          \right)  
	     \right] 
 +\left\lfloor
    \sqrt{\rhoi/\rhoe}\dispfrac{l}{d/R-1}
     + \dispfrac{1}{4}
  \right\rfloor \pi,	     
  \label{eq_step_tmp3}
\end{eqnarray}
   with the floor function $\lfloor \cdot \rfloor$ 
   introduced to make $\phi_l$ continuous with respect to $l$.
Restricting oneself to those modes with $l \gg d/(\pi R)$ and $l \gg kd/\pi$
   in view of the assumptions~\eqref{eq_step_closed_omglargeAssump}, 
   one recognizes that the RHS of Equation~\eqref{eq_step_tmp3} is necessarily far
   smaller than $l \pi$.
Overall, it suffices for our purposes to summarize 
   the properties of the high-frequency modes as
\begin{equation}
\omgl \approx l\pi (\vae/d), \quad 
\mbox{for}~ 
l \gg \dispfrac{d/R}{\pi}, \quad 
l \gg \dispfrac{k d}{\pi}.
\label{eq_app_high_endres}
\end{equation}

\section{Some Further Properties of EVP~1}
\label{sec_app_EVP_highfreq} 
This section examines some further properties of EVP~\ref{evp_mu_closed}
    by capitalizing on the \Schrod\ Equation~\eqref{eq_EVP_schro}.
We recall that the equilibrium density is of the ``outer-$\mu$''
    type (see Equation~\eqref{eq_rho_prof_outermu}).
This subsection extends Subsection~\ref{sec_app_sub_step_modalclosed}
    in that $\mu$ is no longer restricted to be infinite.

We start by showing that all solutions to EVP~\ref{evp_mu_closed} necessarily possess
    an $\omgl$ that exceeds $k\vai$, regardless of $\mu$ or $d/R$.
What we offer is only a slight generalization of
    the arguments given by \citetalias{1996ApJ...472..398B} for a step density profile.
This generalization is possible because the arguments
    therein rely only on two conditions, 
    one being that the potential $V(r)$ consistently exceeds $k^2 \vai^2$,
    and the other being that 
    the eigenfunction $\breve{v}_l(r)$ vanishes at both $r=0$ and $r=d$.
Neither condition is restricted to the particular $\mu = \infty$. 
Now suppose that $\omgl < k \vai$ for some mode, meaning that $\omgl^2 < V(r)$.
The wavefunction $\Phi(r)$ and hence the eigenfunction $\breve{v}_l(r)$ 
    then necessarily peak somewhere in the domain,
    diminishing toward both smaller and larger $r$. 
One then deduces that $d \breve{v}_l(r)/dr$ is discontinuous, 
    thereby violating the continuity requirement for the Eulerian perturbation
    of total pressure.    

We now address high-frequency modes in a system with $d/R \gg 1$, 
   ``high'' in the same sense as in the
    assumption~\eqref{eq_step_closed_omglargeAssump}.
Our approach is essentially a classical WKB one detailed in
    \citet[][Chapter~10]{BenderOrszag1999}.
Somehow different is that we avoid the complication associated with
    the turning points (namely where $V(r) = 0$), which are bound
    to occur at small $r$ 
    for high-frequency modes because $V(r)$ diverges at $r=0$.
This is done by handling the \Schrod\ Equation~\eqref{eq_EVP_schro} 
    in the interior ($r<R$) and exterior ($r>R$) separately. 
First consider the exterior, where the condition
    $\omega^2 \gg V(r)$ is ensured by 
    Equation~\eqref{eq_step_closed_omglargeAssump}.
The leading order WKB solution to Equation~\eqref{eq_EVP_schro}
    reads    
\begin{equation}
\Phi (r) \approx A_{\rm e} \va^{1/2}(r) \sin[\Theta(r)],
	\label{eq_high_PhiWKB}
\end{equation}    
    where
\begin{equation*}
\Theta(r) = \omega \int_{R}^{r}\dispfrac{dr'}{\va(r')}
                +\phi
\end{equation*}
    with $A_{\rm e}$ and $\phi$ being constants.
Requiring $\Phi(d) = 0$ then leads to
\begin{equation}
\omgl \approx \dispfrac{l\pi - \phi_l}{\int_{R}^{d} dr'/\va(r')}
    \label{eq_high_tmp1}
\end{equation}
    for mode $l$.
Evidently, the integral in Equation~\eqref{eq_high_tmp1} is 
    well approximated by $d/\vae$ for large $d/R$. 
The high-frequency modes are therefore
    still characterized by Equation~\eqref{eq_app_high_endres},    
    provided $|\phi_l| \ll l\pi$.

We proceed to show that the inequality $|\phi_l| \ll l\pi$ indeed holds
    via three mutually complementary methods.
Firstly, we numerically solve
    EVP~\ref{evp_mu_closed} for a substantial number
    of combinations of $[\rhoi/\rhoe, \mu; k; d/R]$.
Comparing the computed $\omgl$ with the RHS of 
    Equation~\eqref{eq_high_tmp1} then indicates that 
    $|\phi_l| \ll l\pi$.
Secondly, we make some analytical progress to estimate $\phi_l$.     
Now the interior ($r<R$) needs to be examined, for which purpose
    Equation~\eqref{eq_step_open_vImproper} indicates that
    the exact solution to Equation~\eqref{eq_EVP_schro} reads
    $\Phi(r) = A_{\rm i} r^{1/2} J_1 (\ki r)$ with $A_{\rm i}$ being
    an arbitrary constant.    
In addition, the condition $\ki R \gg 1$ is ensured by
    the assumption~\eqref{eq_step_closed_omglargeAssump}, enabling
    one to employ the approximate expressions for Bessel functions
    at large arguments to find
\begin{equation}
        \left.\dispfrac{d\Phi/dr}{\Phi} \right|_{r = R^{-}} 
\approx \ki \cot\left(\ki R - \pi/4\right)
	    -\dispfrac{1}{2R}.   
    \label{eq_schr_impe_int}
\end{equation}
One further finds with the external WKB solution~\eqref{eq_high_PhiWKB} that
\begin{equation}
        \left.\dispfrac{d\Phi/dr}{\Phi} \right|_{r = R^{+}} 
\approx \ki \cot\phi_l
	   +\left.\dispfrac{1}{2}\dispfrac{d\ln\va}{dr}\right|_{r = R^{+}}.   
    \label{eq_schr_impe_ext}
\end{equation}
Equating Equations~\eqref{eq_schr_impe_int} to \eqref{eq_schr_impe_ext} then 
    leads to that 
\begin{equation}
        \cot\phi_l
\approx 
 		\cot\left(\ki R - \pi/4\right)
 		-\dispfrac{1}{2}
 		 \left(\dispfrac{1}{\ki R}
 		      +\left.\dispfrac{1}{\ki}\dispfrac{d\ln\va}{dr}\right|_{r = R^{+}}
 		 \right).
 		 \label{eq_tmp2}
\end{equation}       
Now suppose that the second term in the RHS of Equation~\eqref{eq_tmp2} is negligible, 
    implying that $\mu$ is not too large.
Let us further see the RHS of Equation~\eqref{eq_tmp2} as known from 
    Equation~\eqref{eq_app_high_endres}.
It turns out that $\phi_l$ can be approximated to leading order by
    $\phi_l \approx \ki R - \pi/4 \approx l\pi \sqrt{\rhoi/\rhoe}/(d/R)$,
    meaning that $|\phi_l| \ll l\pi$ for sufficiently large $d/R$.
Thirdly, we offer some heuristic arguments to estimate $|\phi_l|$
    for arbitrary $\mu>1$. 
Let $N_{\rm int}$ ($N_{\rm ext}$) be the    
	the number of extrema in the eigenfunction $\breve{v}_l (r)$
	in the interior $r<R$ (the exterior $R<r<d$).
Evidently, $\phi_l$ in Equation~\eqref{eq_high_tmp1} 
    stems from the influence of the interior, making
    $|\phi_l/\pi|/l$ essentially a measure of $N_{\rm int}/N_{\rm ext}$.
With $d/R \gg 1$, one readily deduces that $N_{\rm int} \ll N_{\rm ext}$
    and hence $|\phi_l| \ll l \pi$.
     	 
\section{Intricacies Associated with the ``Outer $\mu$'' Formulation}
\label{sec_app_FD_scale}
\begin{figure}
\centering
 \includegraphics[width=.8\columnwidth]{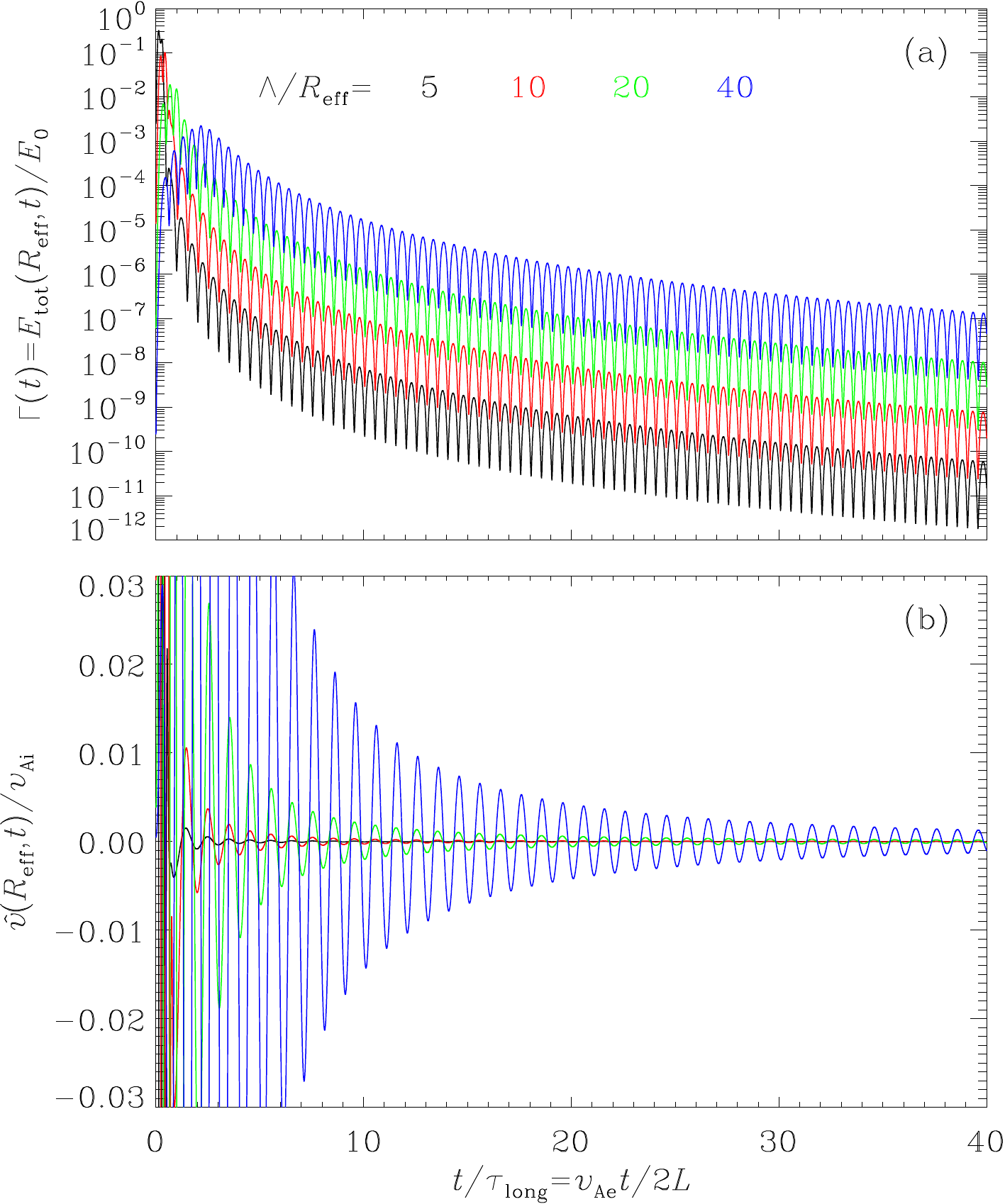}
 \caption{
Temporal profiles for (a) the energy fraction $\Gamma(t)$
   and (b) the radial speed $\hat{v}(\Reff, t)$   
   for a loop with $[\rhoi/\rhoe, L/\Reff] = [10, 10]$.
The steepness parameter is fixed at $\mu = 1.5$,
   while a number of values are examined for the spatial extent of the 
   initial perturbation as labeled. 
All solutions are found with the finite-difference approach.        
Here $\Reff$ represents the effective loop radius, 
   and $\Gamma(t)$ measures the total energy in the volume $r \le \Reff$ 
   in units of the energy imparted to the entire system
   by the initial perturbation.
Note that $\hat{v}(\Reff, t)$ is measured in units of the magnitude
   of the initial velocity perturbation.     
See text for details.    
 }
 \label{fig_app_AR_varLambda}
\end{figure}

{
This section examines some subtlety associated with our
   ``outer $\mu$'' formulation~\eqref{eq_rho_prof_outermu}
   for the equilibrium density $\rho_0(r)$.
Let us recall our argument that, for the 
   the fixed pair $[\rhoi/\rhoe, L/R] = [2.25, 15]$ and some chosen $\mu <2$,
   the dimensionless spatial extent of the initial perturbation ($\Lambda/R$)
   needs to be unrealistically large for evanescent modes to be non-negligible
   in the time-dependent solutions to IVP~\ref{ivp_mu_open}.
However, Equation~\eqref{eq_rho_prof_outermu} indicates that 
   the spatial range containing the density enhancement broadens
   when $\mu$ decreases, making the nominal radius $R$ less and less ideal
   for characterizing the spatial variation of $\rho_0(r)$.
Let the spatial scale of $\rho_0(r)$ be measured by
   some effective radius $\Reff$.
One may question whether our argument still holds because there may be a regime
   where evanescent modes are visible for not so extreme values of
   $\Lambda$ measured in units of $\Reff$ rather than $R$.
Somehow it is non-trivial to quantify this aspect in an exhaustive manner, 
   to explain which we note that we will exclusively adopt
   the FD approach to solve IVP~\ref{ivp_mu_open} here for 
   computational convenience.
We will additionally fix the steepness parameter at $\mu=1.5$, but 
   experiment with two different values for the density contrast $\rhoi/\rhoe$.     
Let $\tau$ denote the duration to be examined in an FD solution, and 
   $D_\tau$ the distance that the outer edge of the perturbation reaches
   when $t = \tau$.   
For now consider the modal structure for EVP~\ref{evp_mu_closed} on a sufficiently
   large domain of size $d$, where by ``sufficiently large" we mean that 
   $d \gg D_\tau$ such that a multitude of evanescent modes exist.
From Section~\ref{sec_modal} we know that oscillatory modes are always permitted,
   and the frequencies of evanescent modes ($\omega < \omgcrit = k\vae$)
   may not differ much from those of the low-frequency oscillatory modes
       ($\omega \gtrsim \omgcrit$).
Now that low-frequency oscillatory modes are excited in general, 
   it may take some considerable amount of time
   for them to interfere such that their contribution to a time-dependent solution
   eventually becomes sufficiently weak to make evanescent modes visible.
In practice, however, the value of $\tau$ cannot be made infinite.
Three regimes then arise in the signal behavior within a large but nonetheless finite
   timeframe, where evanescent modes are not discernible, somehow discernible but weak,
   and stronger than some threshold, respectively.
We see evanescent modes as observationally relevant 
   only when the last regime occurs.
   
Some definitions and remarks are necessary here.
We define $\Reff$ as the radial distance where the function $f(r)$
   in Equation~\eqref{eq_rho_prof_outermu} attains $1/10$, a factor that is 
   meant to be small but admittedly arbitrary. 
One nonetheless finds that $\Reff/R = 10^{1/\mu}$ and evaluates to $4.64$
   for $\mu = 1.5$.
We see $\Reff$ rather than $R$ as being observationally relevant
   and consistently use $\Reff$ to measure $\Lambda$ and the loop length $L$.  
Likewise, time will be measured in units of the longitudinal \Alf\ time
   $\taulongi = 2\pi/\omgcrit = 2L/\vae$, which is more relevant for
   the large-time behavior.
We examine only the timeframe $t\lesssim 40 \taulongi$, which is seen as
   sufficiently long. 
Overall, the time-dependent solutions to IVP~\ref{ivp_mu_open} 
   are determined by 
   the set of dimensionless parameters $[\rhoi/\rhoe, L/\Reff, \Lambda/\Reff]$
   given a fixed $\mu = 1.5$.
We deem the range $\Lambda/\Reff \le 20$ as observationally 
   realistic for initial perturbations.   
On the other hand, we will examine the following two quantities
   to assess the significance of evanescent modes.
The first is $\Gamma(t) = E_{\rm tot}(\Reff, t)/E_0$ 
   with $E_0 = E_{\rm tot}(\Lambda, t=0)$
   being the total energy initially deposited to the entire system.
The second is simply the instantaneous radial speed at the effective radius
   $\hat{v}(\Reff, t)$. 
While both $\Gamma(t)$ and $\hat{v}(\Reff, t)$ measure the signal strength
   in the volume $r \le \Reff$, we find that the former can better bring out
   the differences when $L/\Reff$ or $\Lambda/\Reff$ varies.

We start by examining an AR loop with $[\rhoi/\rhoe, L/\Reff] = [10, 10]$, 
   which is only marginally realistic because in general $\rhoi/\rhoe$ ($L/\Reff$)
   is lower (larger) in observations \citep[e.g.,][]{2004ApJ...600..458A,2007ApJ...662L.119S}. 
Figure~\ref{fig_app_AR_varLambda} displays the temporal profiles of 
   (a) $\Gamma(t)$ and (b) $\hat{v}(\Reff, t)$ for a number of $\Lambda/\Reff$
   as labeled.
Examining any $\Lambda/\Reff$ in the chosen timeframe,
   one sees that a periodic behavior develops  at large $t$ for
   $\Gamma(t)$ and $\hat{v}(\Reff, t)$ alike.
We will focus on this periodic stage here and hereafter.
A slight difference between Figures~\ref{fig_app_AR_varLambda}a and 
   \ref{fig_app_AR_varLambda}b is then that
   the period in $\Gamma(t)$ is half that in $\hat{v}(\Reff, t)$, 
   which arises simply because $E_{\rm tot}$ involves the perturbations 
   as squared terms by definition (see Equation~\eqref{eq_def_ener_Etot}).
More importantly, the signal strengthens when $\Lambda/\Reff$ increases as 
   can be discerned in Figure~\ref{fig_app_AR_varLambda}b
   and seen more clearly in Figure~\ref{fig_app_AR_varLambda}a.
Regardless, the signal for any $\Lambda/\Reff$ weakens monotonically with time,
   eventually resulting in extremely small values for both $\Gamma(t)$
   and $\hat{v}(\Reff, t)$. 
Note that this is true even for $\Lambda/\Reff = 40$, which exceeds the range
   that we deem observationally realistic.
Note further that the signal in the periodic stage tends to weaken when $L/\Reff$
   increases or $\rhoi/\rhoe$ decreases from the value we choose.
Figure~\ref{fig_app_AR_varLambda} therefore means that evanescent modes are not
   discernible for realistic combinations of $[\rhoi/\rhoe, L/\Reff]$, thereby
   strengthening our conclusion that the existence of evanescent modes in the pertinent
   EVP analysis does not guarantee their relevance in the temporal evolution
   of the system. 
It then follows that one needs to invoke, say, kink modes, to account for 
   a periodicity on the order of the longitudinal \Alf\ time when analyzing
   oscillating AR loops even given our outer-$\mu$ formulation.
Furthermore, whether an interpretation in terms of kink modes is justifiable
   can be readily assessed by looking for the tell-tale signature of
   transverse displacements because AR loops tend to be imaged with high
   spatial resolution on a routine basis.

That said, evanescent modes may indeed be discernible or visible if one experiments
   with, say, drastically different values of $\rhoi/\rhoe$.
We proceed with a fixed combination
   $[\rhoi/\rhoe, \Lambda/\Reff] = [100, 20]$, the chosen  density contrast
   being relevant for flare loops \citep[e.g.,][]{2004ApJ...600..458A}.      
Figure~\ref{fig_app_Flare_varL} presents the temporal profiles
   of (a) $\Gamma(t)$ and (b) $\hat{v}(\Reff, t)$ for a variety
   of dimensionless loop length $L/\Reff$ as labeled. 
Three features are evident regarding the periodic stage.
Firstly, Figure~\ref{fig_app_Flare_varL}b indicates that
   the signal for any $L/\Reff$ further settles to a 
   stage where $\hat{v}(\Reff, t)$ actually possesses two periodicities. 
   the dominant one ($P_{\rm domi}$) being close to but nonetheless longer than $\taulongi$.
In addition, the signal amplitude is modulated by a second period 
   ($P_{\rm env} \gg P_{\rm domi}$)
   \footnote{Several full cycles of this periodic amplitude modulation
   can be seen in the time series that extends beyond, say, $120 \taulongi$.
   However, we choose not to show these long time series to avoid the curves becoming too crowded}, 
   as can be seen more clearly in Figure~\ref{fig_app_Flare_varL}a. 
Secondly, when $L/\Reff$ increases, the signal weakens and 
   $P_{\rm domi}$ ($P_{\rm env}$) in units of $\taulongi$
   slightly decreases (increases).
This feature can be seen in both Figures~\ref{fig_app_Flare_varL}a 
   and Figure~\ref{fig_app_Flare_varL}b, but is clearer in the latter.
Thirdly, and more importantly, the signal envelope
   fluctuates about a time-independent level, which can be seen by examining
   the profiles of the maxima/minima of
   either $\Gamma(t)$ or $\hat{v}(\Reff, t)$. 
Given this feature, we take the relevance of evanescent modes
   as self-evident, and focus on the more interesting $L/\Reff$-dependencies
   of $P_{\rm domi}/\taulongi$ and $P_{\rm env}/\taulongi$.
Evidently, the envelope modulation stems from the beat among evanescent modes, which
   themselves possess frequencies that are marginally
   lower than $\omgcrit = 2\pi/\taulongi$.
We recall that a larger $L/\Reff$ means a larger $L/R$ and hence
   a smaller dimensionless axial wavenumber $k R$.
We proceed to consider evanescent eigenmodes of some given radial
   harmonic number $l$
   (say, $l=1$ or $l=2$ as in Figure~7 of \citealt{2017ApJ...836....1Y}).
Let $\omega$ denote the eigenfrequency, $P=2\pi/\omega$ the eigen-period,
   and $\vph = \omega/k$ the phase speed. 
One readily finds that $P/\taulongi = \omgcrit/\omega = \vae/\vph$. 
The reason for $P_{\rm domi}/\taulongi$ to decrease with $L/\Reff$
   is then that the phase speeds of evanescent modes increase toward $\vae$
   when $kR$ decreases 
   \citep[see Figure~7 in][]{2017ApJ...836....1Y}.
On the other hand, that $P_{\rm env}$ increases with $L/\Reff$
   is because a reduction in $kR$ makes smaller
   the difference in the values of $\vph/\vae$ for modes with adjacent values of $l$.
   
That evanescent modes can be seen in Figure~\ref{fig_app_Flare_varL} does not 
   necessarily mean that they are observationally relevant.
Consider only a fixed $\rhoi/\rhoe = 100$ and 
   only the periodic stage.
Let $A$ denote the maximum amplitude of $\hat{v}(\Reff, t)$
   in units of the magnitude of the initial perturbation.
Experimenting with a substantial number of combinations $[L/\Reff, \Lambda/\Reff]$,
   we find that $A$ always decreases (increases) with increasing
   $L/\Reff$ ($\Lambda/\Reff$) when the other parameter is fixed, at least when
   $8 \le L/\Reff \le 15$ and $0 < \Lambda/\Reff \le 20$.
Suppose that evanescent modes can be seen as observationally relevant
   only when $A$ exceeds some threshold $A_{\rm c}$, and let
   $(L/\Reff)_{\rm c}$ denote the critical $L/\Reff$ beyond which
   $A < A_{\rm c}$.
Seeing $A_{\rm c}$ as variable, this then means that $(L/\Reff)_{\rm c}$ 
   is a function of $A_{\rm c}$.
Figure~\ref{fig_app_LRcrit} displays the $A_{\rm c}$-dependence of      
   $(L/\Reff)_{\rm c}$ for a number of $\Lambda/\Reff$ as labeled, before
   describing which we need to mention some technical subtlety. 
We consistently adopt the timeframe $t \le 40 \taulongi$ to determine $A$
   and eventually construct a symbol in Figure~\ref{fig_app_LRcrit}.
However, Figure~\ref{fig_app_Flare_varL}a has already shown that $P_{\rm env}$
   becomes longer when $L/\Reff$ increases, making it possible that
   the amplitude maximum is not captured for $t \le 40 \taulongi$
   if $P_{\rm env}$ is too long.
This turns out not to be a real concern for two reasons, one being that one needs
   to specify a timeframe in any case, and the other being that the envelope modulation
   is already clear for $t \le 40 \taulongi$.
We will return to this point later.
Now let ``allowed range" refer specifically to those $L/\Reff$
   for which evanescent modes are visible.    
One sees from Figure~\ref{fig_app_LRcrit} that 
   $(L/\Reff)_{\rm c}$ decreases monotonically with 
   $A_{\rm c}$ for a given $\Lambda/\Reff$, meaning that the ``allowed range''
   narrows when the relevant instrumental sensitivity weakens. 
Likewise, $(L/\Reff)_{\rm c}$ increases when $\Lambda/\Reff$ increases, meaning that
   the ``allowed range'' broadens with increasing $\Lambda/\Reff$    
   for a given sensitivity and hence a given $A_{\rm c}$. 
   
It should be informative to place Figure~\ref{fig_app_LRcrit} in
   the \citetalias{1983SoPh...88..179E} context. 
Consider an ``ER'' loop, by which we mean a loop with
   an equilibrium density $\rho_0(r)$ that equals $\rhoi$ for $r<\Reff$
   but $\rhoe$ otherwise. 
Restrict ourselves to axial fundamentals.   
A critical $(L/\Reff)_{\rm ER}$ then follows from Equation~\eqref{eq_step_kcut},
\begin{equation}
  \label{eq_tmp3}
  (L/\Reff)_{\rm ER} = \dfrac{\pi \sqrt{\rhoi/\rhoe-1}}{j_{0,1}},
\end{equation}
   only below which evanescent modes are possible in an ER loop. 
With Equation~\eqref{eq_tmp3} one finds that 
   $(L/\Reff)_{\rm ER}$ evaluates to $13$ for $\rhoi/\rhoe=100$, 
   and this value is represented by the horizontal dash-dotted line in 
   Figure~\ref{fig_app_LRcrit}.
Two situations arise regarding the relevance of $(L/\Reff)_{\rm ER}$,
   to describe which we assume that $L/\Reff$ is known observationally
   and that a time series is available for the pertinent observable
   over tens of longitudinal \Alf\ times.  
First consider the situation where $L/\Reff$ is observed
   to exceed $(L/\Reff)_{\rm ER}$.
Suppose that the observed signal eventually possesses a stage where
   the amplitude does not diminish with time.   
It then follows from the left corner of Figure~\ref{fig_app_LRcrit} that
   this stage cannot be interpreted in the \citetalias{1983SoPh...88..179E}
   framework, because no evanescent modes can be excited regardless of $\Lambda$. 
In this regard, a diffuse loop model such as formulated by the outer-$\mu$ profile
   may indeed broaden the range of flare-associated QPPs that
   FSMs may account for. 
However, this broadending is rather limited because it happens only when
   the relevant instrument is highly sensitive and $\Lambda/\Reff$ is nearly
   beyond the range that we deem observationally realistic. 
Now consider the situation where $L/\Reff$ is observed
   to be smaller than $(L/\Reff)_{\rm ER}$.
Evanescent modes are allowed by ER loops in this case,
   and are visible for the majority of combinations $[A_{\rm c}, \Lambda/\Reff]$
   explored in Figure~\ref{fig_app_LRcrit} for our outer-$\mu$ loops. 
One may then question whether the temporal profile of the pertinent observable
   can help categorize the involved flare loop as an ER loop
   or an outer-$\mu$ one.
This is indeed possible, to illustrate which we consider the case where
   $L/\Reff = 11$.
Suppose that the flare loop is describable as an ER loop.
Equation~\eqref{eq_tmp3} then yields a value of $5.66$ if 
   one replaces $j_{0,1} = 2.41$ therein with $j_{0,2} = 5.52$.
This means that the evanescent mode with radial harmonic number $l=2$ is not relevant,
   and the pertinent signal eventually settles to a monochromatic variation
   with a constant amplitude. 
Now suppose that the flare loop agrees with an outer-$\mu$ loop 
   for which $\mu$ is presumed to be $1.5$,
   and suppose further that $\Lambda/\Reff = 10$. 
It turns out that the $\hat{v}(\Reff, t)$ profile is very similar to 
   the black curve in Figure~\ref{fig_app_Flare_varL}b, even though
   the amplitude maximum $A$ in the periodic stage attains a smaller
   value of $\sim 0.1$. 
Nonetheless, it an amplitude of this magnitude can be resolved,
   then the relevant observable will be seen to experience
   some envelope modulation, which is distinct from what is expected for
   an ER loop.
In addition, this amplitude modulation may also be useful to distinguish
   between kink modes and FSMs in flare loops, if flare loops are not 
   well spatially resolved to tell whether they are experiencing transverse displacements.
We refrain from discussing this aspect further for two reasons, 
   one being that kink modes in our outer-$\mu$ setup have yet to be examined,
   and the other being that a dedicated forward modeling approach seems necessary to 
   establish the detailed observational signatures. 
Rather, with Figure~\ref{fig_app_LRcrit} we conclude that whether evanescent modes
   in outer-$\mu$ loops can be observed depends critically on instrumental
   sensitivity.
We conclude further that the \citetalias{1983SoPh...88..179E} cutoff
   $(L/\Reff)_{\rm ER}$ serves as a useful reference 
   in that evanescent modes in outer-$\mu$ loops 
   are hardly observable when $L/\Reff \ge (L/\Reff)_{\rm ER}$
   unless the pertinent instrument is extremely sensitive.    
}

\clearpage

\clearpage

\clearpage

\clearpage

\clearpage

\clearpage
\clearpage

\clearpage
\clearpage

\clearpage

\clearpage

\clearpage

\clearpage
\begin{figure}
\centering
 \includegraphics[width=.8\columnwidth]{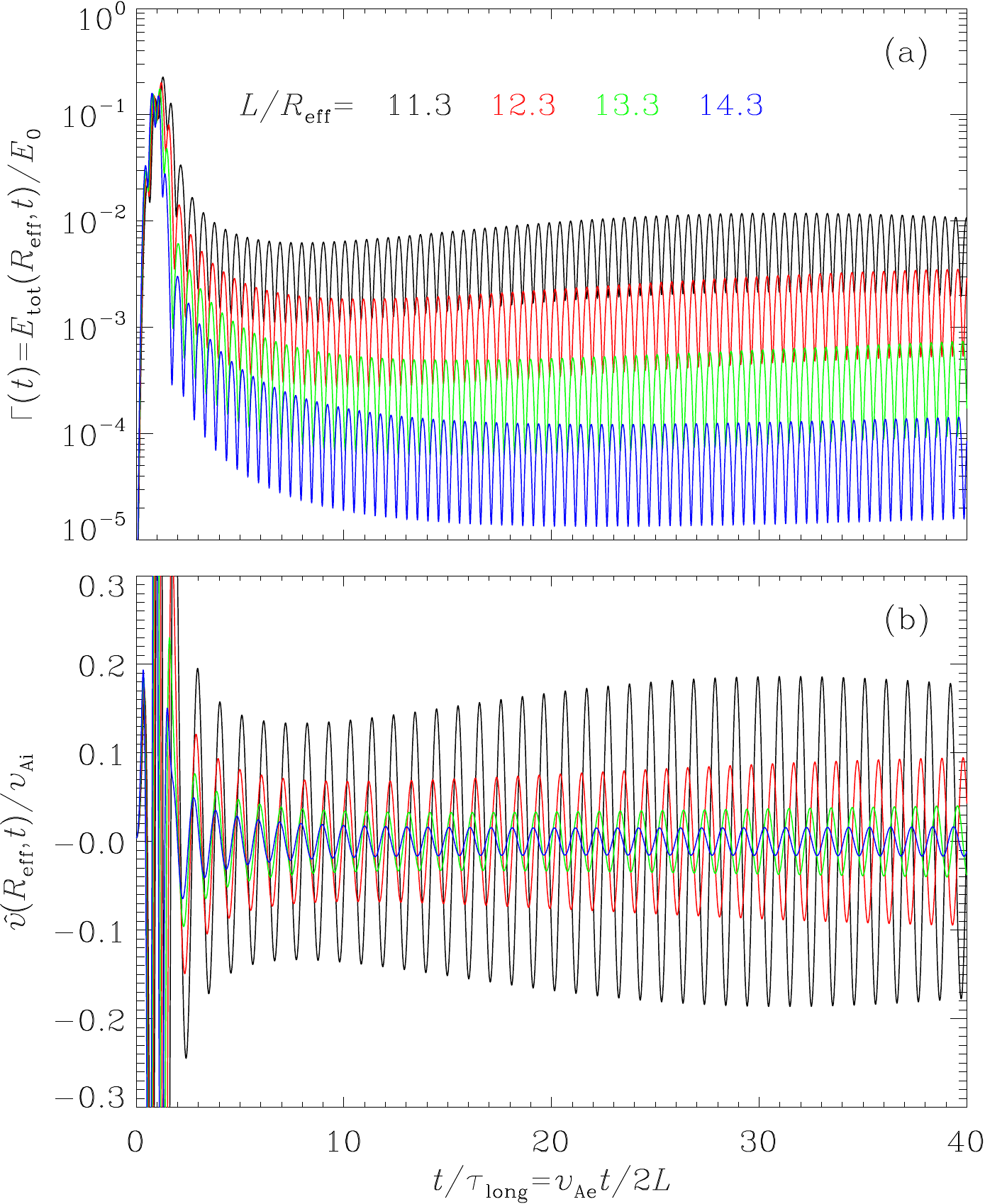}
 \caption{
Similar to Figure~\ref{fig_app_AR_varLambda} but for 
   $[\rhoi/\rhoe, \Lambda/\Reff] = [100, 20]$.
The steepness parameter is fixed at $\mu = 1.5$,
   while a number of values are examined for $L/\Reff$ as labeled. 
See text for details.    
 }
 \label{fig_app_Flare_varL}
\end{figure}

\clearpage
\begin{figure}
\centering
 \includegraphics[width=.8\columnwidth]{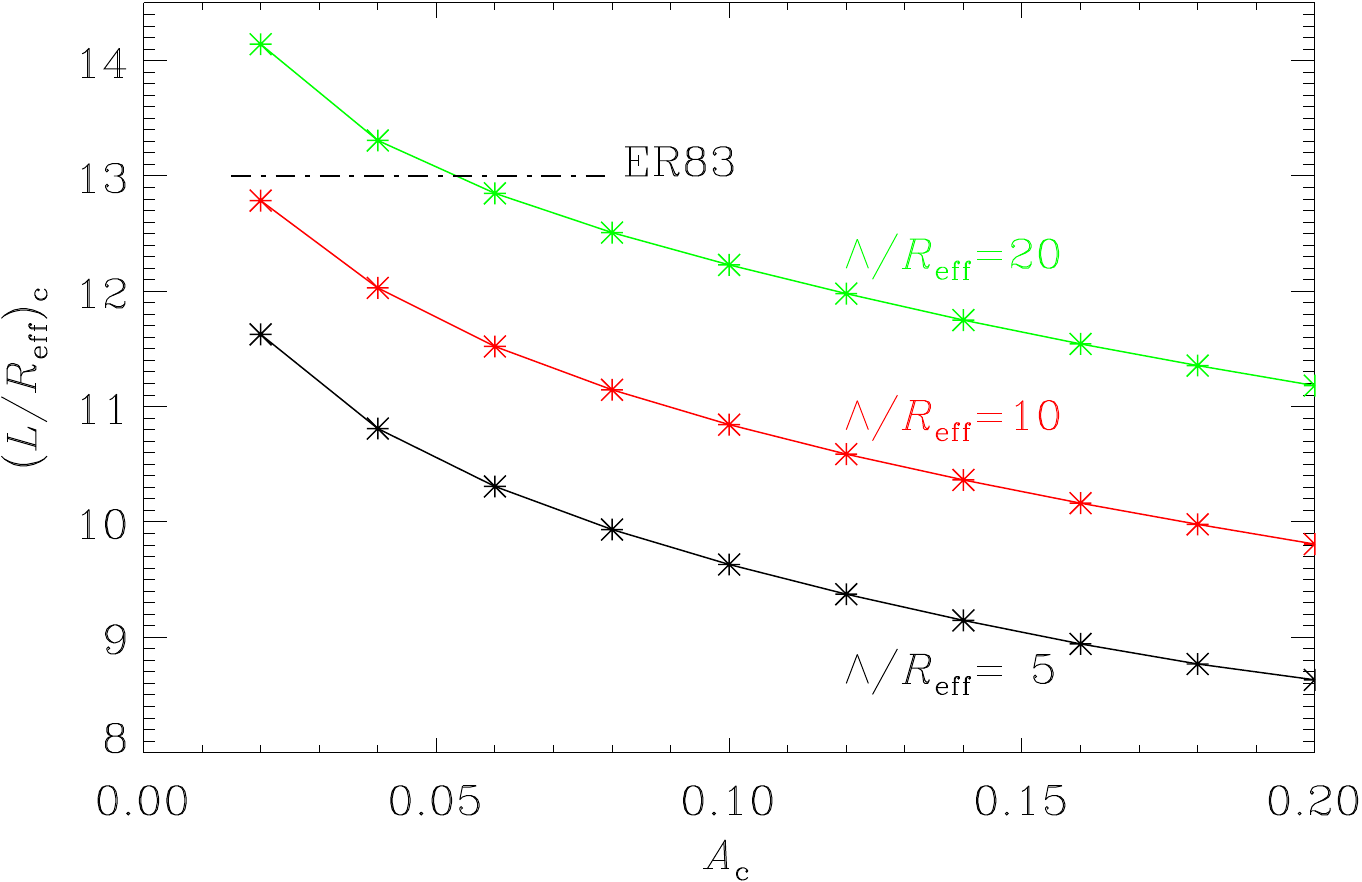}
 \caption{
Dependence of the critical $(L/\Reff)_{\rm c}$ on the critical dimensionless
   amplitude $A_{\rm c}$ for a number of $\Lambda/\Reff$ as labeled. 
The combination $[\rhoi/\rhoe, \mu]$ is fixed at $[100, 1.5]$.    
For a given $\Lambda/\Reff$, a loop with $L/\Reff$ larger (smaller) than  
   $(L/\Reff)_{\rm c}$
   yields a $\hat{v}(\Reff, t)$ for which the maximum amplitude
   in the periodic stage is smaller (larger) than $A_{\rm c}$ when measured 
   in units of the magnitude of the initial velocity perturbation. 
The horizontal dash-dotted line represents the expectation within
  the \citetalias{1983SoPh...88..179E} framework 
  ($(L/\Reff)_{\rm ER}$) for a piece-wise uniform
  loop where the equilibrium density attains $\rhoi$ for $r\le \Reff$
  but $\rhoe$ otherwise. 
Evanescent modes are not relevant when $L/\Reff < (L/\Reff)_{\rm ER}$ 
  in this framework. 
See text for details.    
 }
 \label{fig_app_LRcrit}
\end{figure}

\end{document}